\documentclass{IEEEoj}
\usepackage{cite}
\usepackage{amsmath,amssymb,amsfonts}
\usepackage{algorithmic}
\usepackage{graphicx,color}
\usepackage{textcomp}

\usepackage{orcidlink}
\usepackage{comment}
\usepackage{float}  

\def\BibTeX{{\rm B\kern-.05em{\sc i\kern-.025em b}\kern-.08em
    T\kern-.1667em\lower.7ex\hbox{E}\kern-.125emX}}
\def\OJlogo{\vspace{-10pt}\includegraphics[height=20pt]{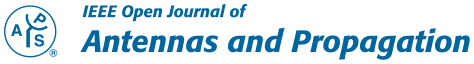}}
\begin{document}

\title{Beam-pattern Prediction for Radio Antenna Arrays using Mutual Coupling and Machine Learning}

\author{Anna M. Heckel\authorrefmark{1}\orcidlink{0009-0009-1613-8331},
Chris J. Pate\authorrefmark{1}\orcidlink{0009-0006-9779-1057},
Jungmin Kim\authorrefmark{2}\orcidlink{0000-0001-8567-5338}, 
Peter T. Timbie\authorrefmark{1}\orcidlink{0000-0003-0325-1633}, and
Zongfu Yu\authorrefmark{2}\orcidlink{0000-0002-4536-1526},~\IEEEmembership{Fellow,~IEEE}}
\affil{Department of Physics, University of Wisconsin, Madison, WI 53706 USA}
\affil{Department of Electrical and Computer Engineering, University of Wisconsin, Madison, WI 53706 USA}
\corresp{: Peter T. Timbie (e-mail: pttimbie@wisc.edu).}

\begin{abstract}
Radio antennas are often used to form arrays, such as phased arrays and interferometer arrays.  For most applications the beam patterns of these instruments must be accurately known.  For phased arrays the array patterns are formed by coherently adding the embedded element patterns (EEPs) of the individual antennas in the array. For interferometers, the fundamental measurement is of visibilities, which depend on the products of pairs of EEPs.  Unfortunately, in either case, for closely-spaced arrays, the EEPs can be distorted by mutual coupling between the antennas and can be very different from the \emph{isolated} element patterns (IEPs), measured when the antennas are far apart.  Mutual coupling is characterized by the scattering parameters for the array.   Full-wave electromagnetic simulations can simulate the EEPs, including mutual coupling effects, but these simulations can be computationally expensive and require accurate knowledge of the design of the array elements, their placement, and the environment in which they are deployed. Directly measuring either the EEPs or the array patterns for instruments in the field is often impractical as well. In this paper we show that in cases where the scattering parameters between array elements can be measured, they can be used to infer the array patterns of a phased array.  We demonstrate the training of a deep neural network to simulate array patterns when given: the design of the antennas, the locations of the antennas, the frequency of operation, and measurements of the scattering parameters of the array.   As a  demonstration for a simple model, we apply this approach to the simulation of the array patterns for a small phased array of three dipole antennas and evaluate the sensitivity of this procedure to uncertainties in measurements of the antenna positions and scattering parameters.
\end{abstract}



\maketitle

\noindent\textit{Preprint submitted to IEEE Open Journal of Antennas and Propagation.}

\section{INTRODUCTION}
\IEEEPARstart{A}{rrays} of radio antennas are ubiquitous in applications such as communications, radar, and radio astronomy. Radio interferometer arrays perform aperture synthesis to create high resolution images of the sky, while phased arrays create array beams that have higher directivity than can their antenna elements, form multiple beams simultaneously, and rapidly steer them.  However, designing and testing most arrays is challenging.   One must know the radiation patterns of each array element {\it when embedded in the array} (i.e. the embedded element patterns, or EEPs) to predict the beam patterns of the full array. Unfortunately, when closely spaced, the array antenna elements interact with each other electromagnetically through mutual coupling (MC);  the received or transmitted radio waves scatter off the other antennas in the array.  When antennas are embedded in an array, MC distorts their radiation patterns away from their isolated element patterns (IEPs).  In some applications, such as 21\,cm intensity mapping in radio astronomy, small uncertainties in the EEPs can lead to systematic errors that dwarf the small expected cosmic signals \cite{Di_Ninni_mutual_2022,Rath2025,Ohara2025}.

Array beams can be computed using full-wave electromagnetic simulation codes that account for MC effects even in arrays with hundreds of complicated antennas in close proximity \cite{Bolli2023,Gueuning2022, Gueuning2025}. However, the actual patterns will be sensitive to details of the as-built antennas, such as their exact shapes, material properties, relative positions, and their electromagnetic environment (e.g. presence of the ground, trees, etc.).  Beam patterns of actual arrays can be measured by transmitting signals from calibrated radiation sources at antenna test facilities or by flying transmitters mounted on drones over an array in the field \cite{Tyndall2025}, but these measurements are time-consuming and must be repeated if the array characteristics change with time.  
  
MC between array elements can be fully described by treating the array as a microwave network characterized by its scattering (S) parameters or, equivalently, its mutual impedance matrix (MIM).  Crucially, the S-parameters are related to the EEPs of the array elements \cite{Aumann1989, Craeye2011,Warnick2018,Buck2023}. Indeed, \cite{Kyriakou2025} demonstrate the determination of the MIM of a large antenna array using measurements of the EEPs using the technique of \cite{Buck2023}.  In this paper we perform the reverse operation, to determine the beam patterns of a phased array from knowledge of the S-parameters.  Indirectly, we are also determining the EEPs, which are combined to create the array beam. (Note that MC can also arise from cross-talk between cables, after the antennas, but we assume those effects can be made negligible with appropriate shielding, etc.)

The relationship between EEPs and the S-parameters does not, in general, fully constrain the EEPs, and thus the array beams.  The number of parameters required to specify the antenna patterns is generally larger than the number of S-parameters, at least for arrays with a finite number of elements.  Nevertheless, useful constraints can be placed on the array beams from knowledge of the S-parameters.  In this paper, we demonstrate through simulation how machine learning (ML) can be used to determine array beams, given the S-parameters and other information about an array.  That this might be feasible in principle is supported by the work \cite{Dash2025}, who show that the effect of MC on the EEPs of an array can be described with a relatively small number of parameters.

While antenna patterns are difficult to measure rapidly in the field, S-parameters for arrays are typically not measured at all.  In principle, they can be measured using a vector network analyzer (VNA) \cite{DiLullo_2023,Kwak2024}, but for an array with a large number of elements or that is physically large, such a measurement would be time consuming.  Furthermore, the antenna ports in the array are often not readily accessible.  However, the Appendix proposes the concept of an ``embedded VNA" incorporated into each receiver or transmitter, which would allow the array to automatically measure its own S-parameters by transmitting calibration signals toward, and receiving them from, the other antennas in the array.  

ML has been used to improve antenna behavior prediction, reduce the number of simulations required, and speed up the antenna design process \cite{Podder2023}. Onat el al. \cite{Onat2025} demonstrated that neural networks (NNs) can be trained on suites of full-wave simulations to efficiently simulate EEPs in the presence of MC.  An example of using S-parameter measurements to infer antenna patterns is given by Tamminen et al. \cite{Tamminen2021}, who showed that a fully-connected NN can be trained to map a frequency-swept $S_{11}$ parameter to an antenna beam pattern. This work focused on a single antenna with varying beam widths and steering angles, which faced a quasi-random reflective mask. The parameter $S_{11}$ was calculated as the coupling coefficient between this antenna and the reflected field, which was fed into a NN. The outputs are the beam direction in x ($\theta_x$), beam direction in y ($\theta_y$), beamwidth in x ($\Delta\theta_x$), and beamwidth in y ($\Delta\theta_y$).  By focusing on beam direction and width from frequency-dependent reflection measurements distorted by a random reflector, Tamminen et al.\cite{Tamminen2021} avoided performing full radiation pattern measurements. The direction error is given as $4.1^{\circ}$ and the width error is $2.2^{\circ}$, making this a coarse characterization rather than a high-resolution pattern prediction. 

In this paper, we go beyond this single-antenna case and demonstrate a ML approach to determine the array beams of small antenna arrays. These were chosen to be as simple as possible to allow rapid assessment of the technique, rather than for a particular application.  Specifically, we chose compact arrays of three coplanar, wire dipole antennas with uncertain positions. We present a deep neural network (DNN) that takes in a frequency, the positions of three dipoles which form a small antenna array, and the array's simulated $\bf S$-matrix, and returns a full predicted far-field radiation pattern.

In \S\ref{sec:method} we review the mathematical formalism for defining the EEP's and for combining them to form array beams.  We also describe the simulation of array beams and S-parameters that are then used to train a DNN.  In \S\ref{sec:results} we show how we validated and tested the DNN for three different cases:  A - accurate antenna positions and S-parameters are provided to the trained DNN; B - erroneous antenna positions but accurate S-parameters are provided; and C - accurate antenna positions but inaccurate S-parameters are provided.  In \S\ref{sec:discuss} we quantify the accuracy of the array patterns inferred by the DNN and in \S\ref{sec:conc} we speculate on future directions this approach may take.

\section{METHODOLOGY}
\label{sec:method}

Following Craeye et al. \cite{Craeye2011}, the electric field in the far-field of antenna element $n$ embedded in an array of antennas is given as: 

\begin{equation}
\vec{E}_n(R,\hat{u})=\vec{f}^e_n(\hat{u})\frac{e^{-jkR}}{R}v_{\circ,n},
\label{eq:E}
\end{equation}
where $\hat{u}$ is the direction of $\vec{E}_n$, $R$ is the distance from the antenna, and $k_{n}$ is the wavenumber of the radiation. 
The EEP for antenna $n$ is $\vec{f}^e_n(\hat{u},\nu)$, the far-field pattern of the antenna,  defined 
assuming a Thevenin equivalent source with unit excitation voltage $(\nu_{\circ,n} = 1)$ and series
impedance $Z_{L,n}$,  while the other antenna elements are passively terminated.   The antennas are assumed to be single-mode. 

 Craeye et al. \cite{Craeye2011} give the relationship between the $\bf S$ matrix and the matrix $\bf B$ defining the Beam Coupling Factors of the EEPs for all pairs of antennas in an array.  Specifically, they provide the beam overlap integral (their Eq. 8), which defines $b_{m,n}$, the components of $\bf B$ for antenna elements $m$ and $n$:

\begin{equation}
b_{m,n}=\frac{1}{2\eta}\int_{\Omega} \vec{f}^e_m(\hat{u},\nu)\cdot\vec{f}^{e*}_n(\hat{u},\nu)d\Omega,
\label{eq:B}
\end{equation}
where the integration is over the unit sphere.   $\eta$ is the impedance of free space. 
For the case of lossless antennas, $\bf B $ is related to the scattering matrix, $\bf S $, for the antenna network:

\begin{equation}
{\bf B} = \frac{1}{8 \mathrm{Re}\{Z_L\}}(\bf{I-S^T S^*}),
\end{equation}
where $Z_L$ is the input impedance of the receiver or transmitter circuit connected to the antennas (assumed here to be the same for all ports), ${\bf I}$ is the identity matrix,  and $\bf S^T$ and $\bf S^*$ are the transpose and complex conjugate of $\bf S$, respectively.  Because $\bf B$ is related to the EEPs through an integral relation (Eq. \eqref{eq:B}), fully recovering the EEPs from $\bf S$ is not possible in general. However, as shown below, the relation provides a strong constraint on the EEPs.

Once the EEPs are known, the array patterns for a phased array of $N$ elements with uniform excitation can be determined \cite{Mailloux2018} as follows:

\begin{equation}
\vec{E}(\hat{u})=\sum_{n=1}^{N} \vec{f}^e_n(\hat{u})e^{j k \hat{u}\cdot\vec{r}_n},
\label{eq:Earray}
\end{equation}
where $\vec{r}_n$ is the position vector of antenna $n$. (In general, the excitations can be weighted non-uniformly, but we chose to test the case of uniform excitation here for simplicity.)

Similarly, for the case of an interferometric array, the EEPs can be used to interpret visibility measurements.  Visibilities are the fundamental observable from an interferometer. They depend on the product of the EEPs of pairs of antennas, $m$ and $n$, embedded in an array:  

\begin{equation} V_{mn} = \frac{k_B}{\lambda^2} \int_{4\pi} T(\hat{u})\, \vec{f}^{\,e}_{m}(\hat{u}) \cdot \vec{f}^{\,e*}_{n}(\hat{u})\, e^{-j k \hat{u}\cdot(\mathbf{r}_m-\mathbf{r}_n)} \, d\Omega , 
\label{eq:vis}
\end{equation}
where $k_B$ is Boltzmann's constant, $\lambda$ is the wavelength, $T(\hat{u})$ is the brightness distribution across the sky and $\vec{r}_n$ and $\vec{r}_m$ are the position vectors of antennas $n$ and $m$. We leave the study of interferometers to future work.

In order to simulate EEPs that are consistent with both the S-parameter measurements and our knowledge of the antenna design and environment we use a supervised machine learning approach. 
The motivation for this approach is that while S-parameters do not uniquely determine full radiation patterns, a DNN can learn to infer radiation patterns from the antenna positions, operating frequency, and complex S-parameters. Once trained, the model can predict radiation patterns for previously unseen antenna configurations without requiring additional electromagnetic simulations. The training data set for the neural network is created with full-wave electromagnetic simulations of the S-parameters and array patterns of simple antenna arrays. The data are simulated for arrays of identical antennas over a variety of antenna locations.    
The fidelity of the inferred antenna patterns depends on how finely the parameters are explored in the simulations, so the training hinges critically on  performing the simulations rapidly.  We used a recently-developed commercial tool, called Tidy3D \cite{Tidy}, a cloud-based simulation solver that models interactions between EM waves and structures using the finite-difference time-domain (FDTD) method.  

NNs are trained by adjusting their parameters to minimize a loss function in an iterative manner.  In our case the loss function is based on the difference between the parameters that describe the ``Predicted" beam pattern for the array and the ``Actual" beam pattern.   The Predicted beam is the one chosen by the NN.  The Actual beam is the one simulated by Tidy3D, which we take as truth.  As a loss function we use the mean squared error (MSE) between the Predicted and Actual beam parameters.  During training we iteratively adjust the weights for the network and propagate forward until we converge on the set of weights that minimizes the MSE.
\subsection{Simulations}
\label{sec:sims}
\begin{figure}
    \centering
    \includegraphics[width=0.9\linewidth]{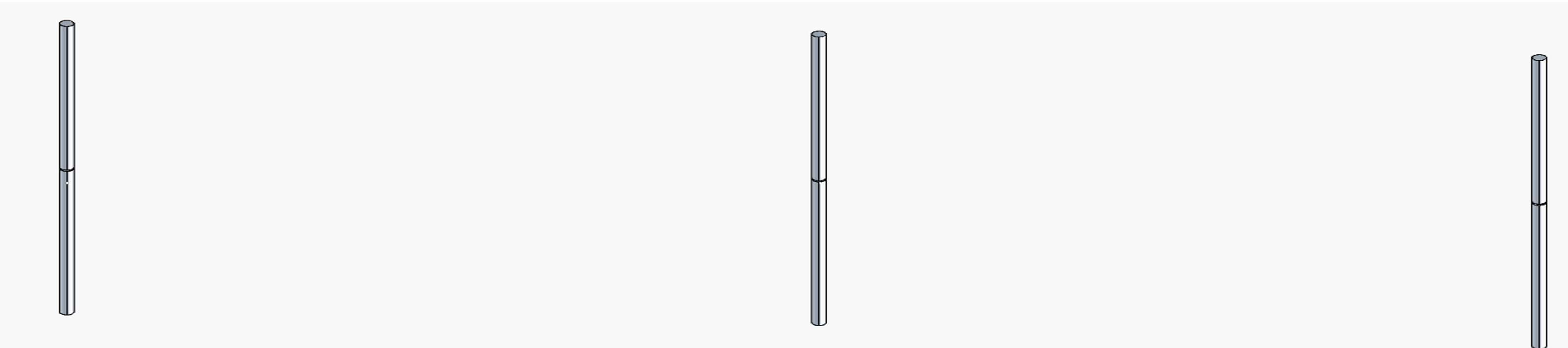}
    \caption{One of the 1000 array configurations modeled using Tidy3D. Each array includes three identical center-fed dipole antennas.}
    \label{fig:3dips}
\end{figure}

\begin{figure}
    \centering
    \includegraphics[width=0.8\linewidth]{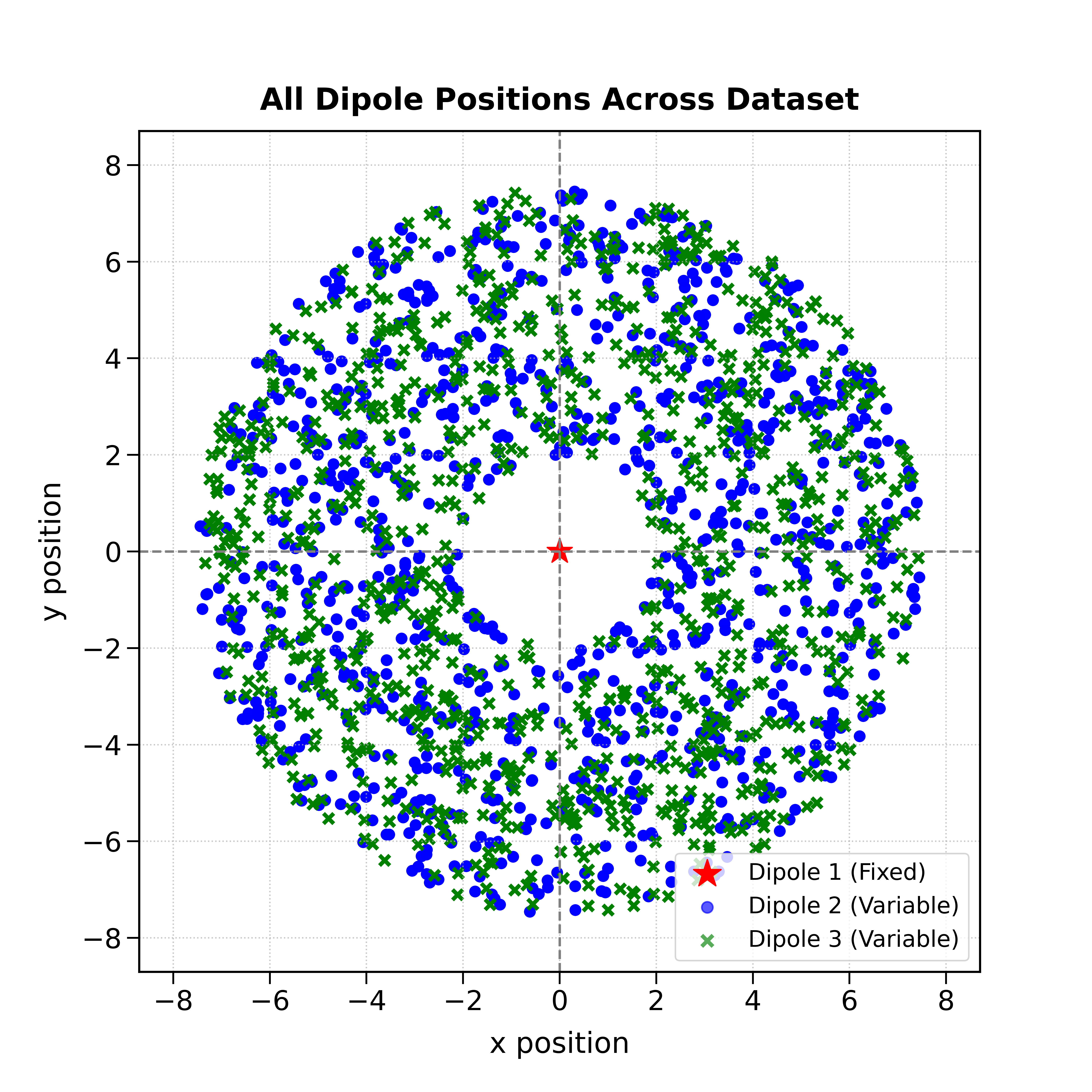}
    \caption{This display of the dipole locations in the x-y plane shows the range of array configurations used for the training set. The dipole axes are oriented along the z axis (out of the page) and positions are in cm.}
    \label{fig:donut}
\end{figure}

Using Tidy3D, we simulated the beam patterns and S-parameters for 1000 configurations of a simple antenna array, made of three dipole antennas.   One such array is shown in Fig. \ref{fig:3dips}. The dipole antennas were modeled to be identical thin wires with an AC source at a port at their centers. The full length of the dipoles was approximately half a wavelength. The antennas were all oriented vertically along the z-axis, with their centers placed at $z=0$. The dipole positions in the xy plane were randomly generated, with constraints: one dipole fixed at the origin, with the other two dipoles located within 0.2 and 0.75 times the wavelength ($\sim 2$ cm and $7.5$ cm, respectively) of the origin, and a minimum separation between the two moving dipoles themselves of 0.18 times the wavelength ($\sim 1.8$ cm). These constraints prevented any antenna configuration from having the dipoles too close to one another, while also ensuring strong interactions (MC) between the antennas and a manageable simulation domain. These restrictions created a doughnut-shaped region in which the two moving dipoles could be placed, as shown more clearly in Fig. \ref{fig:donut}.

We created the training data as follows.  For each of the 1000 array configurations, simulations were performed at five frequencies (2.6, 2.8, 3, 3.2, and 3.4 GHz), which resulted in 5000 total sets of input data. Each of these sets included the x and y-coordinates of the moving dipoles (with the z-position excluded since it was held constant), the frequency, a matrix of S-parameters simulated by Tidy3D (9 real and 9 imaginary numbers), yielding 23 input parameters per set. The EEP for each dipole was simulated by exciting each one (the ``active antenna") sequentially for each configuration while the other two antennas were passively terminated. The data were returned to a unique HDF5 file.  The array's collective far-field beam pattern is determined by summing the electric field contribution from all three dipoles (in phase) for every $\theta$ and $\phi$  to get $E_{\theta}$ and $E_{\phi}$. The directivity is obtained with Eq. \ref{eq:directivity}:
\begin{equation}
    D(\theta, \phi) = \frac{4\pi \left( \lvert E_{\theta} \rvert^{2} + \lvert E_{\phi} \rvert^{2} \right)}{\displaystyle \int_{0}^{2\pi} \int_{0}^{\pi} \left( \lvert E_{\theta} \rvert^{2} + \lvert E_{\phi} \rvert^{2} \right) \sin\theta \, d\theta \, d\phi}.
    \label{eq:directivity}
\end{equation}
A plot of this directivity in 3D space  represents the far-field radiation pattern of each configuration of the full, 3-antenna array, operated as a phased array, with each antenna in phase with the others.

The final data set is a combination of the input parameters, the simulated beam pattern for each set of inputs, and array configuration identification numbers (IDs), ranging from 0 to 999. The configuration IDs are used to identify the 1000 different configurations of the antenna positions. When splitting the data into sub-training, validation, and test blocks, we used these IDs to prevent the model from seeing the same antenna positions in testing as it had in training by placing all simulations with the same ID in the same block. 
In total there were 5000 rows of data, one for each simulated beam pattern.  Five beam patterns for each configuration took about 3 minutes on a high-end workstation with a Graphics Processing Unit. (See Table \ref{tab:dataset} for summary.) 
\begin{table}[H]
\centering
\caption{Dataset Summary}
\label{tab:dataset}
\begin{tabular}{|c|c|c|}
\hline
Content & Shape \\ \hline
Model Inputs & \\ 
(coordinates, frequency, S-matrix) & [5000x23] \\ \hline
Simulated Beam Pattern & [5000x65341] \\ \hline
Simulation ID's & [5000x1] \\ \hline
\end{tabular}
\end{table}

\subsection{Training the DNN}
\label{sec:training}
Prior to training the DNN, 
we split the data set of 1000 array configurations (5000 beams) into training and test sets. We then ensured that no pre-existing, non-physical pattern existed due to the order of the simulation data by randomizing the training data order while keeping array configurations together. Additionally, we further split the training data into a sub-training block and a validation block. The sub-training block contains data used to fit the model and optimize its parameters, while the validation block contains data withheld from the fitting to assess the model's ability to generalize to unseen data. This ensures the model does not use the test dataset as validation data during training, so it sees the test data only during testing.

The next processing step involved reducing the dimensionality of the beam patterns using Principal Component Analysis (PCA). Our Actual (simulated) beam patterns initially consisted of one data point for every combination of $\theta$ and $\phi$ in one degree increments. There are 181 $\theta$ points and 361 $\phi$ points, resulting in $181 \times 361 = 65341$ directivity values per beam pattern. This output dimensionality is substantially larger than the 23 input parameters (1 frequency, 4 coordinates for the moving dipoles, and 9 real and 9 imaginary S-parameter values), making direct prediction of the full beam patterns challenging with the available dataset. PCA constructs a projection matrix that identifies those directions (principal components) that capture the largest variance in the beam patterns. We tested 50, 70, 90, and 200 components in a hyperparameter grid search and found that 70 provided the best balance between dimensionality reduction and preservation of beam structure. (See Table \ref{tab:PCAcomponentsRMSE}). During training, the DNN minimizes the mean squared error (MSE) between the Predicted and Actual PCA coefficients. The Predicted coefficients are then transformed through inverse PCA, which reconstructs the 65,341-value beam pattern as a linear combination of the corresponding principal component basis vectors learned from the training data. These reconstructed beams are compared with the Actual beams in \S\ref{sec:results}.

\begin{table}
\centering
\caption{Hyperparameter grid search results: Comparison of DNN root mean square error (RMSE, in dBi), varying only the number of PCA components.}
\label{tab:PCAcomponentsRMSE}
\begin{tabular}{|c|c|c|c|c|}
\hline
\# PCA components & 50 & 70 & 90 & 200\\ \hline
RMSE & 1.97 & 1.89 & 1.99 & 2.19\\ \hline
\end{tabular}
\end{table}

Finally, we implemented standard scale factors to ensure that the parameters with larger values were not interpreted as more important by the neural network. Otherwise, values such as frequency, which are on the order of $10^9$, would have been automatically weighted higher than values such as the coordinates, which are on the order of 1.

In our initial DNN models, we used a structure similar to the DNN used in \cite{Mehta2019}. The goal is to map a highly non-linear continuous function (positions, frequencies, S-parameters $\Rightarrow$ beam patterns) and we thus require many linear regions given the number of parameters. A DNN provides sufficient layers to learn and accurately model the connections between these parameters, including smaller local changes, such as sidelobes. The network can also fold the input space to  map similar array layouts to similar beam patterns more efficiently.

However, as we added more layers to increase complexity, we observed an increase in the MSE between the Actual and Predicted patterns, suggesting our DNN was overfitting or experiencing vanishing gradients. These phenomena can be common in deeper models trained on small datasets such as ours. Therefore, within our DNN, we constructed residual blocks. These enable the network to learn corrections to the relationships between our inputs and outputs, rather than relearn entirely new relationships. They contain the standard linear layer and the ReLU\cite{Agarap_2026} activation function as well as batch normalization and dropout. Batch normalization was added because our model is large, and gradients can become unstable. Dropout was added to mitigate overfitting. The final model we used was therefore a fully connected deep residual neural network.

\begin{table}
\centering
\caption{Hyperparameter Values}
\label{tab:hyperparams}
\begin{tabular}{|c|c|}
\hline
Number of PCA components & 70 \\ \hline
Train/validation/test split & 700/100/200 \\ \hline
Hidden layers & 7 \\ \hline
Number of Residual blocks & 3 \\ \hline
Layer dimensions & Input: 23$\rightarrow$256  \\
& Blocks: 256$\rightarrow$256, 256$\rightarrow$256, \\ & 256$\rightarrow$128 \\
& Output: 128$\rightarrow$70 \\ \hline
Learning rate & 0.0005 (+ scheduler) \\ \hline
Dropout probability & 0.05 \\ \hline
Weight decay & 0.00001\\ \hline
Batch size & 64 \\ \hline
Maximum training epochs & 300\\ \hline
\end{tabular}
\end{table}

\begin{figure}
    \centering
    \includegraphics[width=0.9\linewidth]{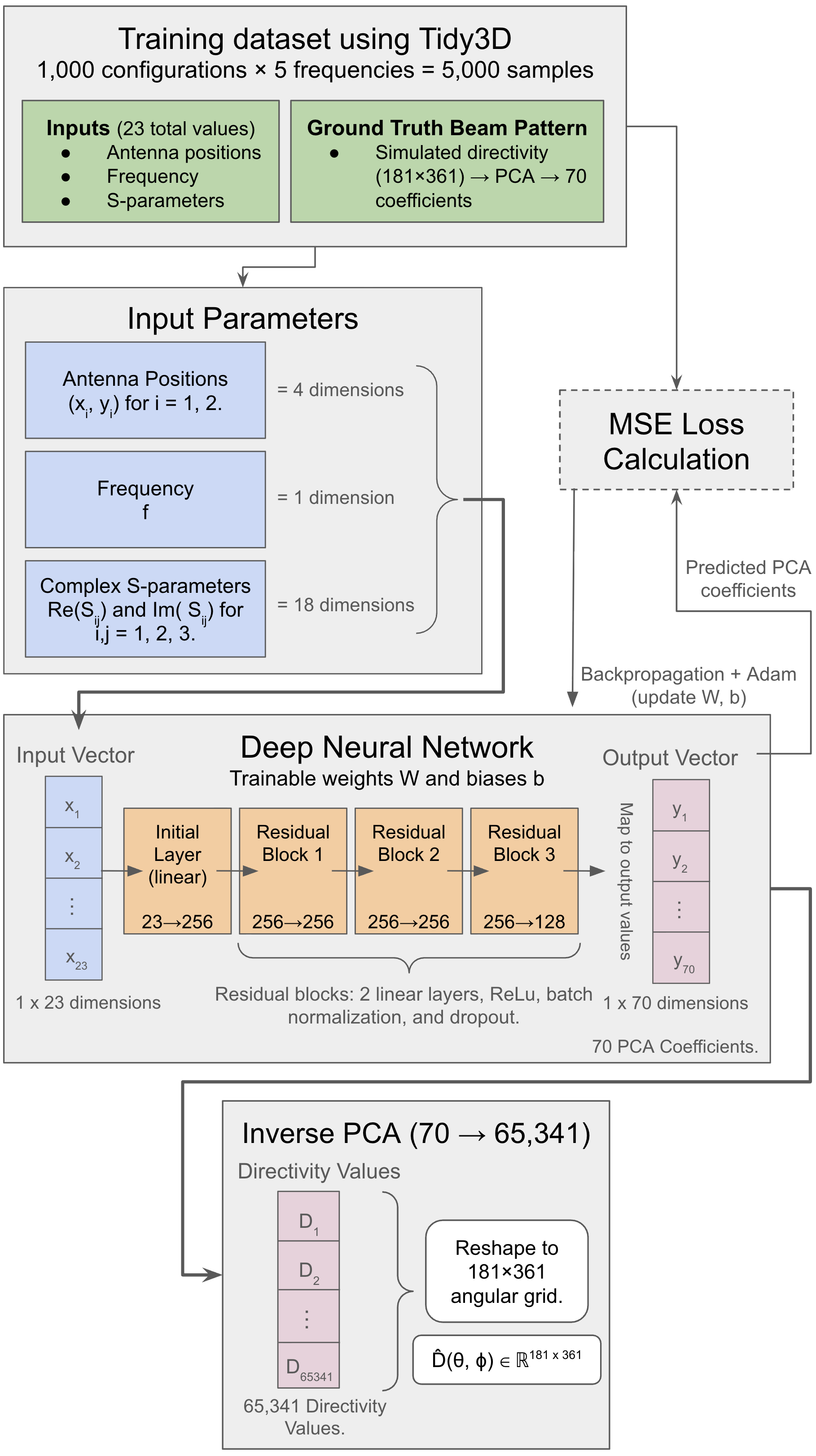}
    \caption{A block diagram of the supervised learning framework used to predict far-field radiation patterns of three-element antenna arrays.}
    \label{fig:dnn_block_diagram}
\end{figure}

The final step of the implementation of the model was to refine its hyperparameters, summarized in Table \ref{tab:hyperparams}. The number of PCA components determines the dimensionality of the compressed beam representation predicted by the DNN. The train/validation/test split specifies the number of samples used to fit the model, tune hyperparameters, and test the model, respectively. The number of hidden layers, residual blocks, and layer dimensions determine the depth and size of the network. Learning rate controls the step size during optimization, with a scheduler used to adjust this value throughout training. Dropout and weight decay are regularization techniques used to minimize overfitting, with dropout randomly deactivating a fraction of neurons during training and weight decay penalizing large model weights. The batch size specifies the number of training samples processed before each optimization step, while the number of epochs defines the maximum number of complete passes through the data set. We systematically optimized the number of PCA components, learning rate, dropout probability, and weight decay through a grid search. However, doing so for all of the hyperparameters would have been computationally expensive. The remaining hyperparameters were selected based on preliminary results and computational considerations.

For model training, we used the standard loop: compute the Predicted beam, compute the loss, update the weights, and then propagate forward again. We chose Adam\cite{Kingma_2017} as our optimizer because it is the standard choice and works for a wide range of problems. We used the MSE as our loss function because we want larger errors to be more heavily penalized than smaller ones to ensure the dominant beam structure is predicted. We also used a scheduler from the Torch\cite{Paszke_2019} package, which allowed fine-tuning of beams towards the end of training. The best weights found during training were saved and applied to the test dataset to obtain our final results. The overall architecture and training procedure of the proposed supervised learning framework are illustrated in Fig. \ref{fig:dnn_block_diagram}.

\section{Results}
\label{sec:results}
We validated and tested the DNN for three cases.  In case A, we determined the MSE between the Actual and Predicted array patterns for the situation where both the S-parameters and the correct antenna positions are provided to the DNN.  Effectively, we used the DNN as a substitute for a full-wave electromagnetic simulator (such as Tidy3D) to find the beam patterns.  In case B, we mimicked a hypothetical situation where the array antennas are deployed with \emph {inaccurate} positions but the S-parameters are still measured accurately. This case tests whether the DNN can account for uncertainties in the deployed array, and hence be more useful than a full-wave electromagnetic simulator. Once again, we used the MSE between the Actual and Predicted array patterns as a figure of merit.  Finally, in case C, we mimicked a situation where the antenna positions are known accurately, but the S-parameters are not.

\begin{table}[H]
\centering
\caption{RMSE (in dBi) for the sub-training, validation, and test splits for each case after reconstructing the full beam from PCA.}
\label{tab:CasesRMSE}
\begin{tabular}{|c|c|c|c|}
\hline
Case & Subtrain & Validation & Test \\ \hline
A & 1.37 & 1.89 & 1.89 \\ \hline
B & 1.37 & 1.88 & 2.09 \\ \hline
C & 1.37 & 1.89 & 1.97 \\ \hline
\end{tabular}
\end{table}

\subsection{Correct antenna positions and S-parameters}
\label{sec:correct}

The RMSE for the sub-training, validation, and test data sets for case A can be seen in Tab. \ref{tab:CasesRMSE}. The difference between the subtrain RMSE vs the validation and test data sets indicates some potential overfitting to the sub-train data. The strong similarity between the validation and test split RMSE values suggests that the model generalizes consistently to new data and recovers the correct array beam patterns with an RMSE of 1.89 dBi.


\begin{figure} 
    \centering
    \includegraphics[width=0.40\textwidth]{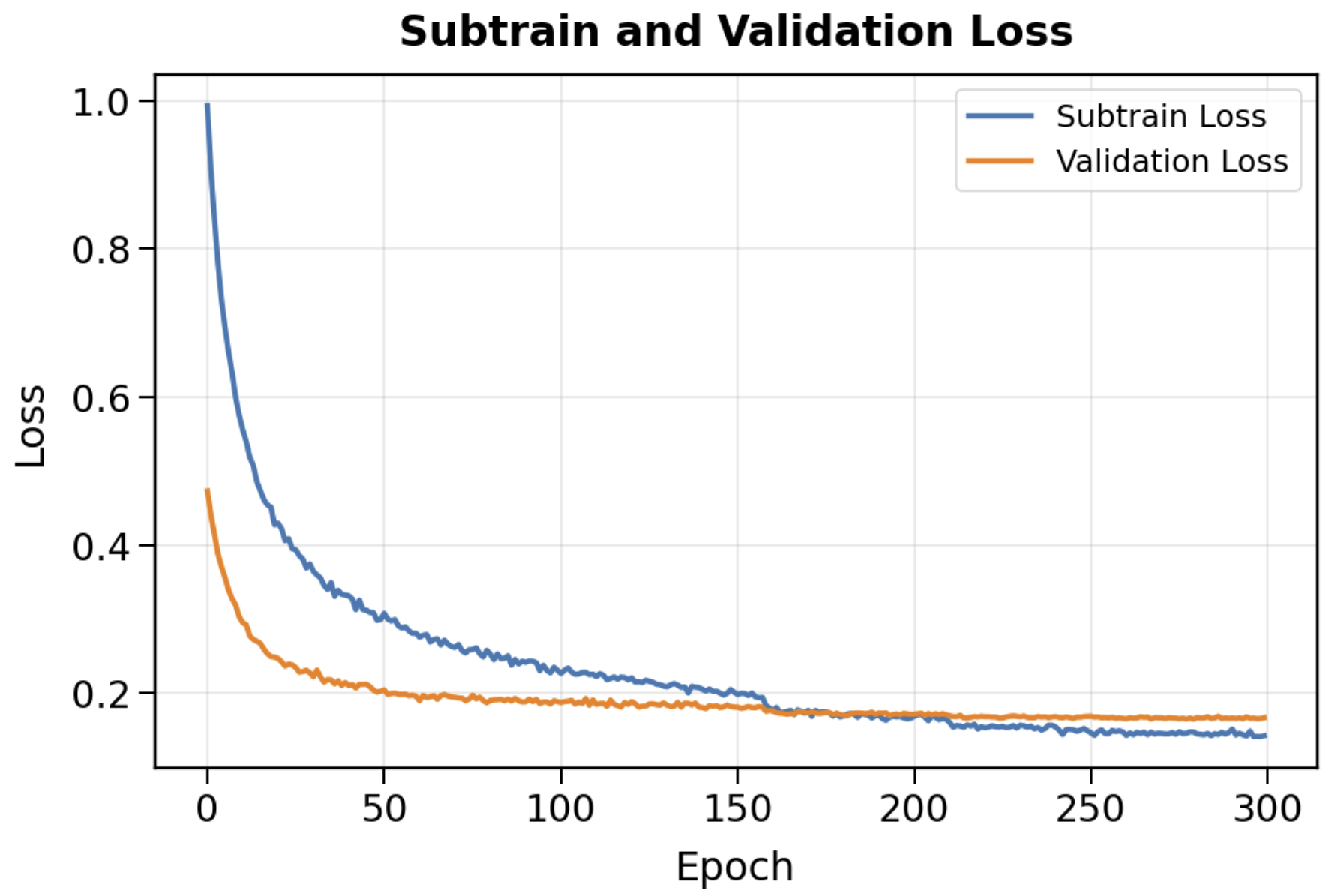}
    \caption{Subtrain and validation loss over 300 epochs for case A.}
    \label{fig:loss}
\end{figure}

\begin{figure}
    \centering
    \includegraphics[width=0.35\textwidth]{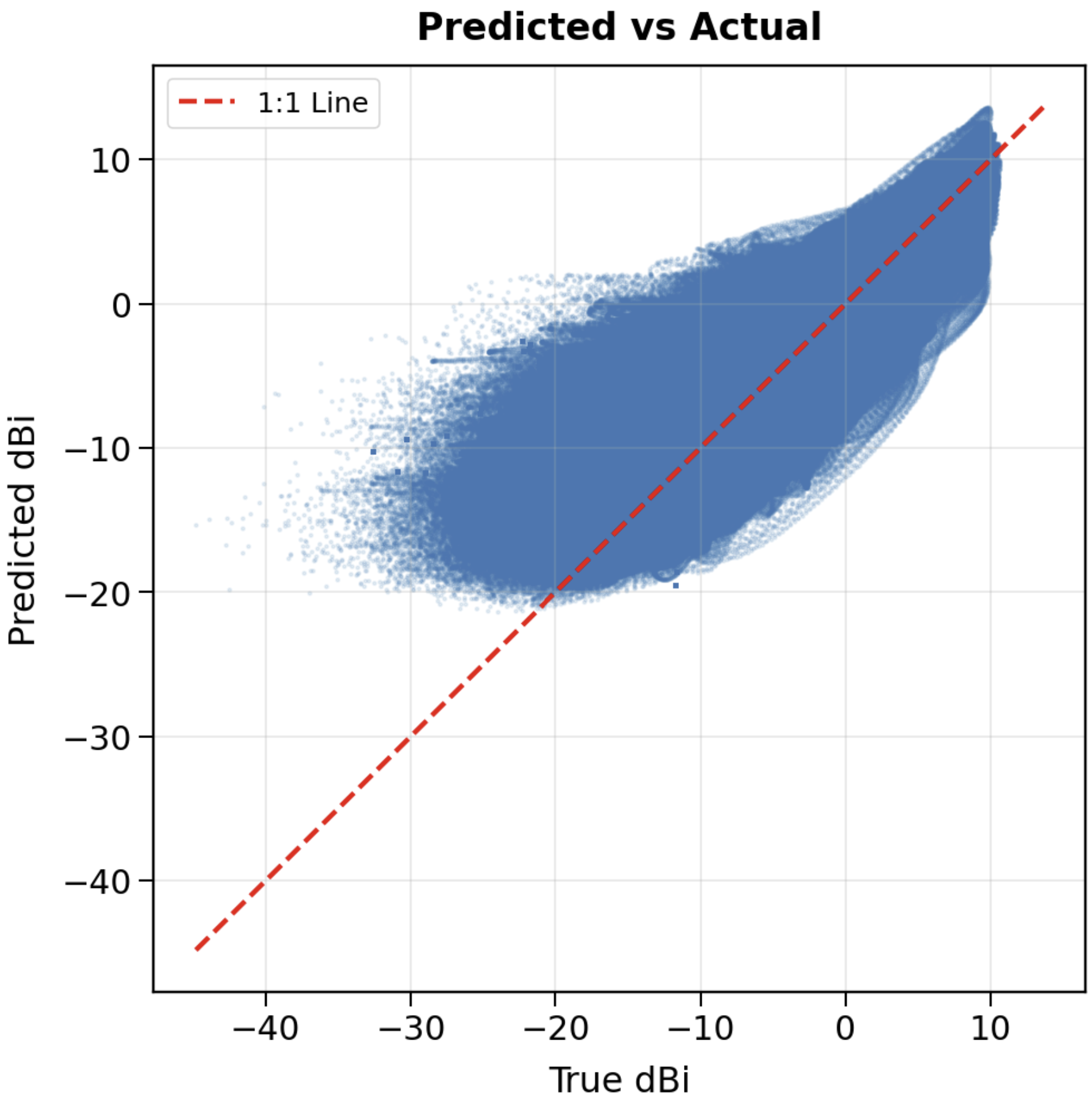}
    \caption{Predicted vs Actual beam directivity (in dBi) for all points in the tested beams ($1000\times65341\sim6.5\times10^7$ points) for case A. The standard deviation (RMSE) of this distribution is 1.89 dBi.}
    \label{fig:pred_v_act}
\end{figure}

\begin{figure}
    \centering
    \includegraphics[width=0.4\textwidth]{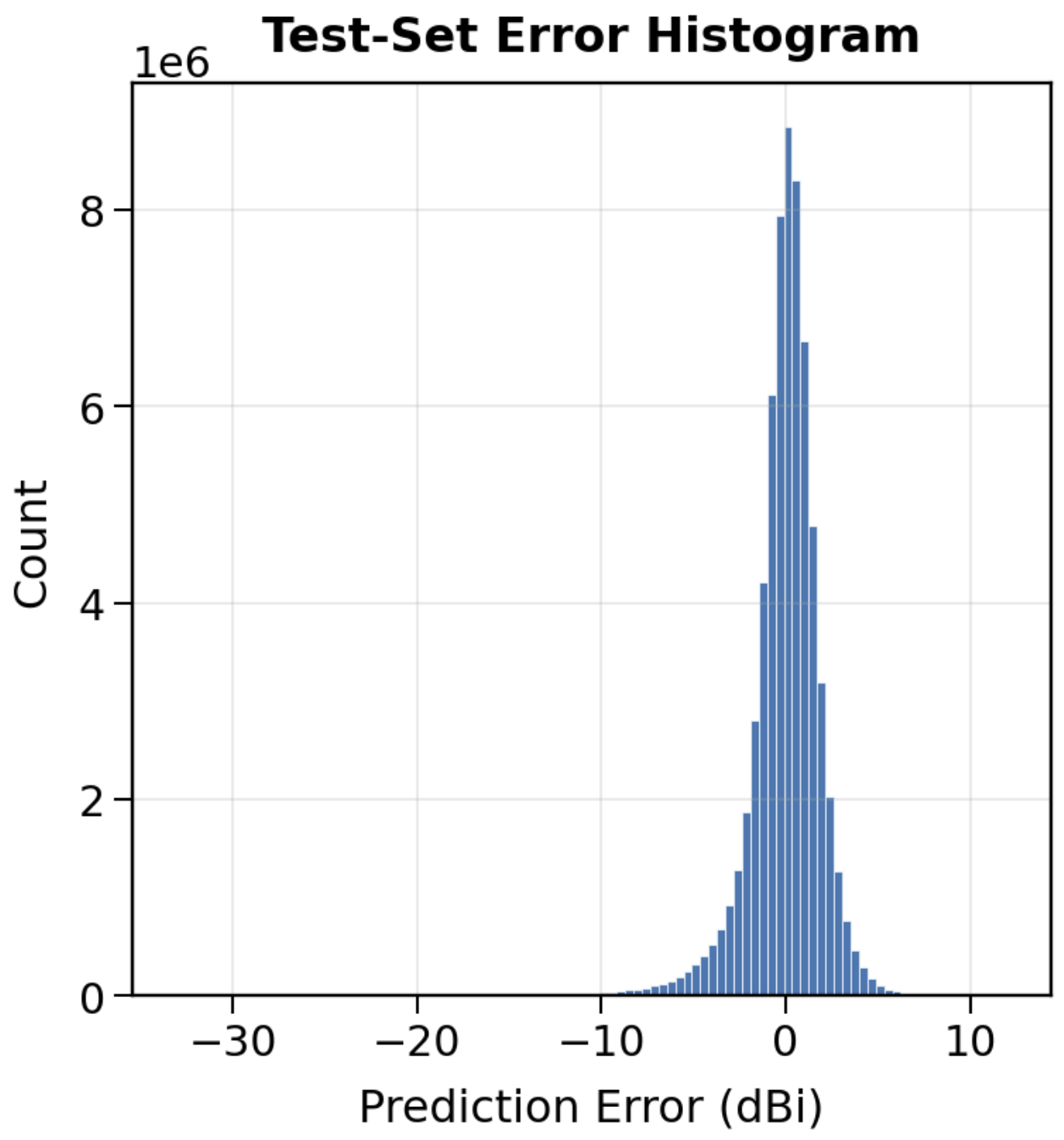}
    \caption{Histogram of the prediction error distribution for the same $\sim6.5\times10^7$ points for case A. The prediction error is defined as the difference between the actual dBi and the predicted dBi of a point in the beam.}
    \label{fig:err_hist}
\end{figure}

The subtrain and validation loss over the course of training (300 epochs) can be seen in Fig. \ref{fig:loss}. These losses converge as the model is trained, suggesting minimal overfitting bias. The discrepancy between the validation and training losses in earlier epochs is likely due to the structure of the DNN. To avoid near-zero weights when increasing the number of layers, the introduction of additional noise in the training data via residual connections (ResNets) was necessary \cite{Shen2017}. It was therefore more difficult to accurately predict the beams using the noisier training data than the validation data, at least in early training. 


The parity plot in Fig. \ref{fig:pred_v_act} displays the Predicted vs. Actual beam directivity in dBi for the Predicted values for all the test cases. The model is more likely to overestimate directivity values than underestimate, and appears to particularly struggle with very small values (below -20 dBi). This is reasonable considering that the structure of the DNN discourages the prediction of extreme values.  Similarly, Fig. \ref{fig:err_hist} shows the distribution of error values for all of the tested cases.

\begin{figure}
\centering 
\makebox[0.33\textwidth][l]{\small (a)}
\includegraphics[width=0.33\textwidth]{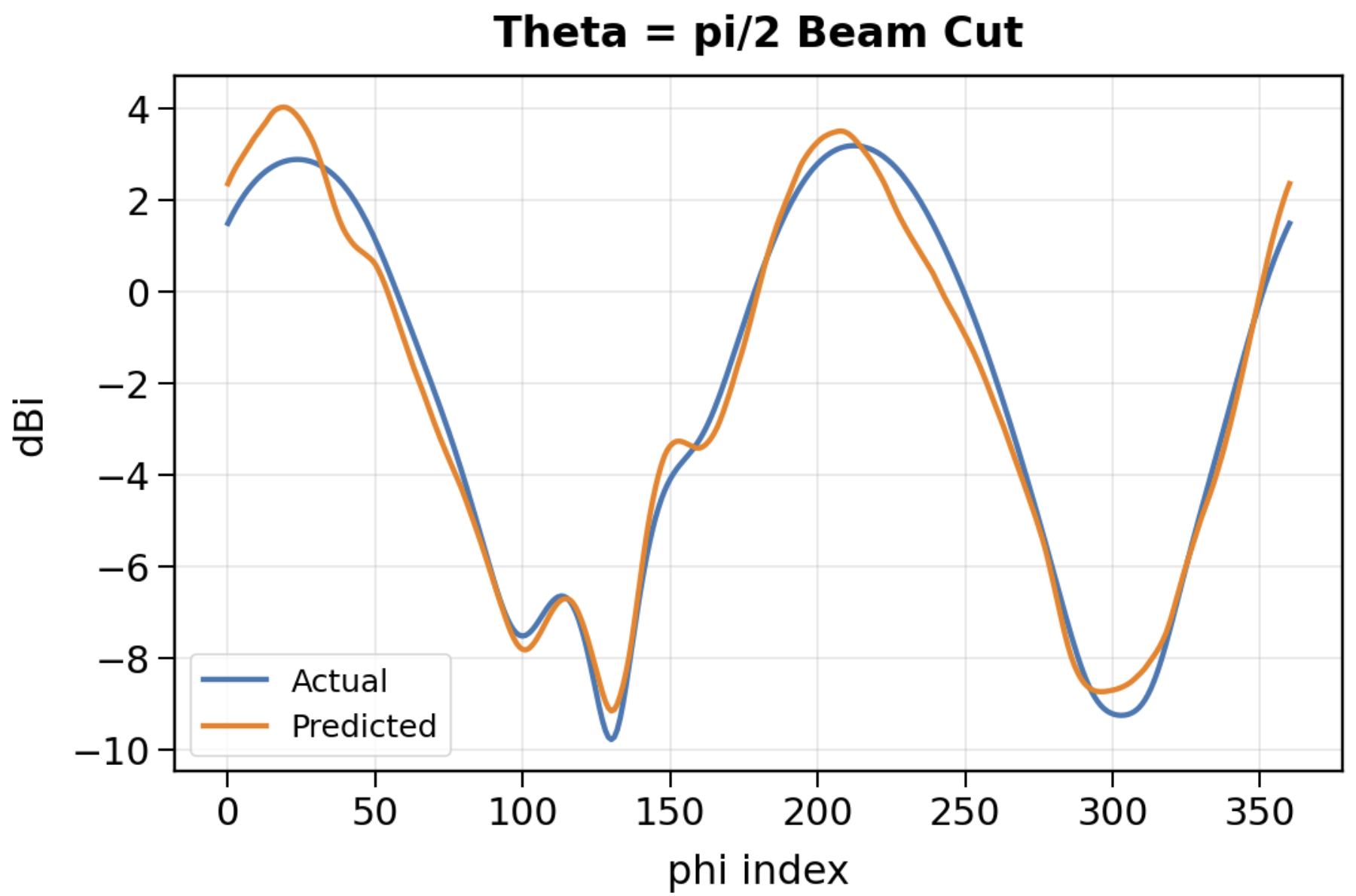}
\makebox[0.33\textwidth][l]{\small (b)}
\includegraphics[width=0.33\textwidth]{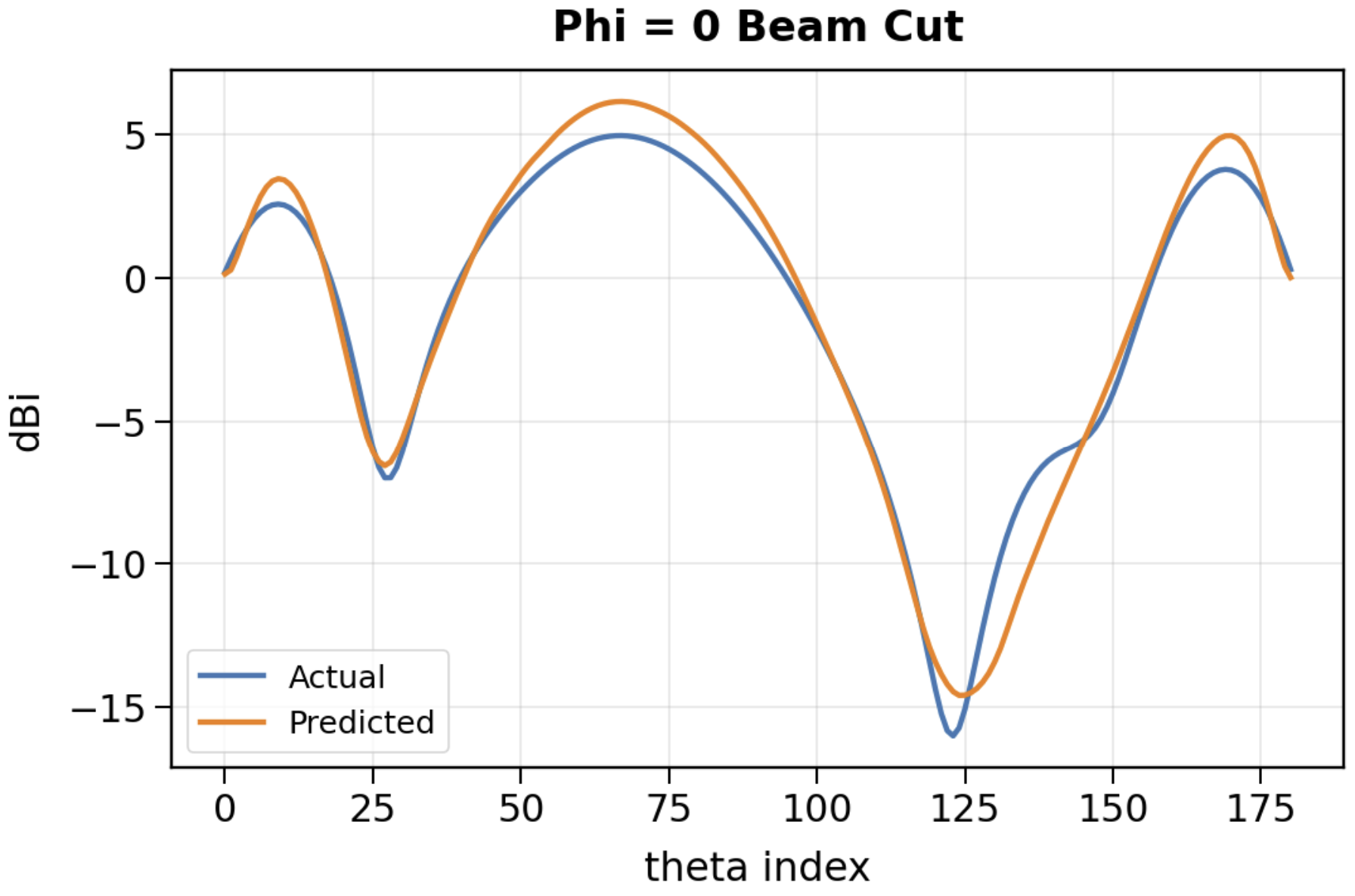} 
\makebox[0.33\textwidth][l]{\small (c)}
\includegraphics[width=0.20\textwidth]{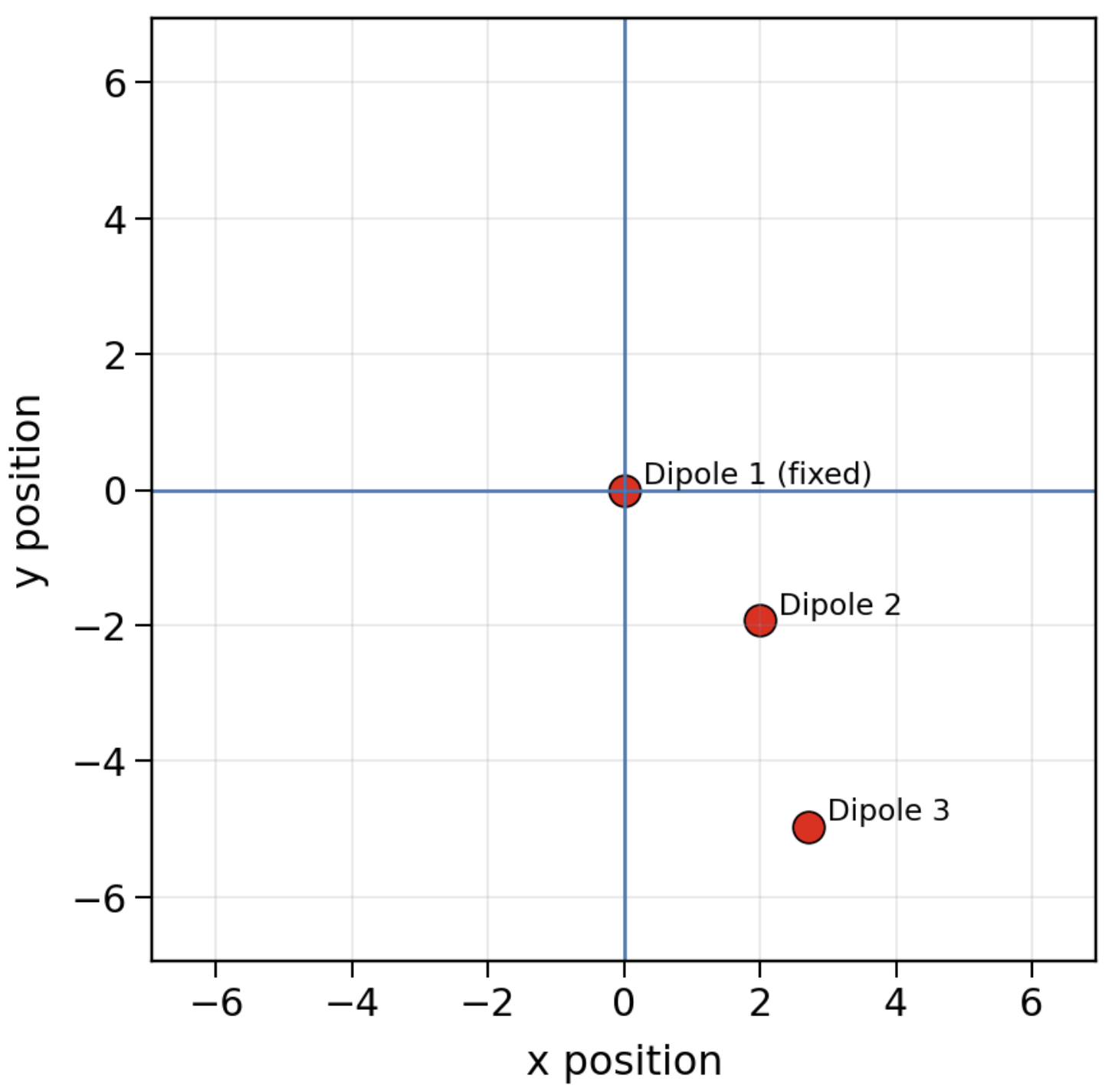}
\includegraphics[width=0.48\textwidth]{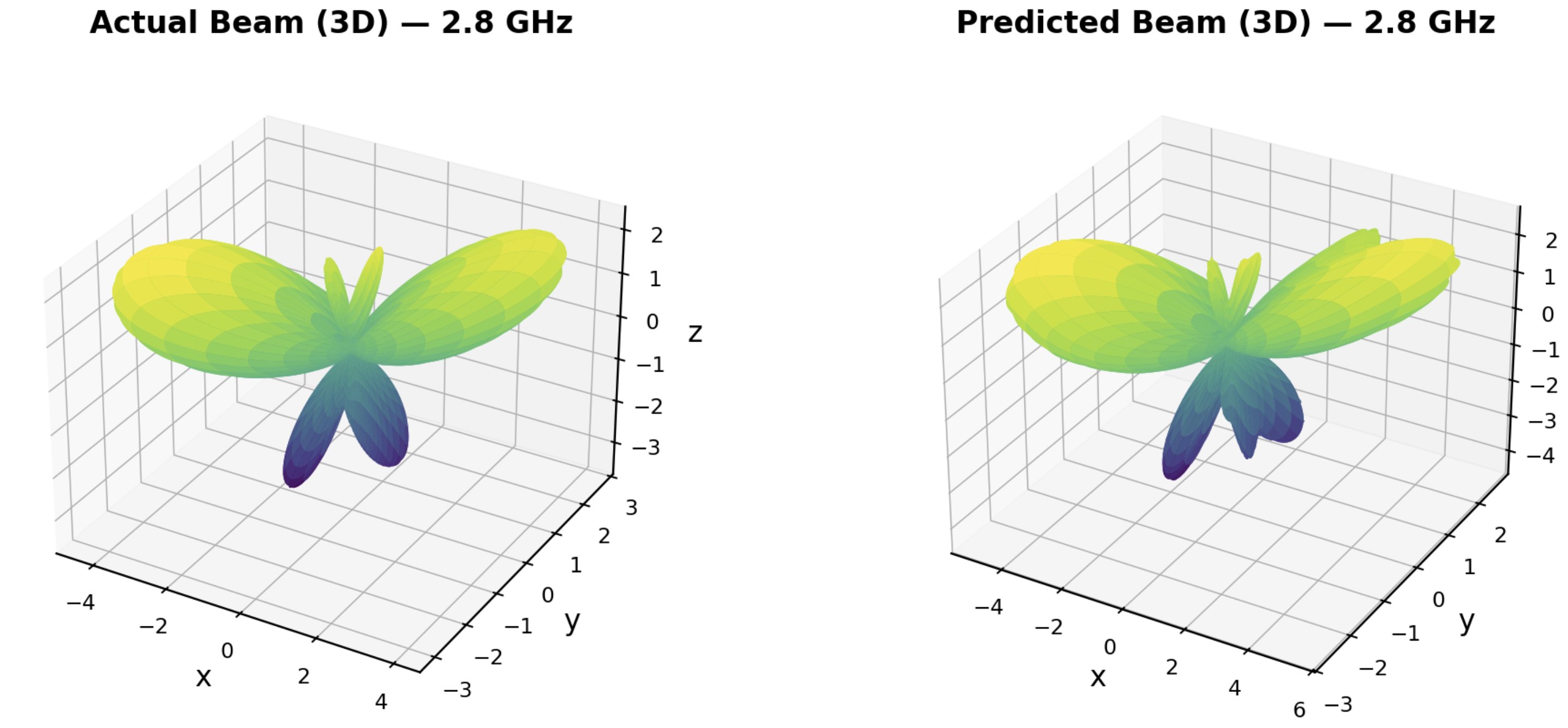} 
{\small (d)}
\caption{Array beam patterns for one particular 3-element array that had the best agreement between the beam pattern predicted by the neural network and the actual beam (simulated by Tidy3D) for case A. The beam patterns are functions of the usual three-dimensional polar coordinates $\theta$ and $\phi$, with $\theta = 0$ oriented along the $z$-axis. Panel (a) shows the $\theta=\pi/2$ cut, panel (b) shows the $\phi=0$ cut, panel (c) shows the antenna positions, and panel (d) compares the Actual (left) and Predicted (right) 3D beam patterns. Axis units are dBi.} 
\label{fig:best_MSE} 
\end{figure}

Fig. \ref{fig:best_MSE} displays the $\theta=0$ and $\phi = \frac{\pi}{2}$ beam cuts from the 3-antenna array that had the best agreement (i.e. the lowest RMSE) between the beam pattern Predicted by the DNN and the Actual beam simulated by Tidy3D. The close agreement between the Predicted (orange line) and Actual (blue line) beams illustrates the DNN's ability to predict beam patterns over a wide dynamic range in power and across a wide range of angles $\phi$ and $\theta$. Fig. \ref{fig:best_MSE}(d) depicts the full 3D beam from which these cuts were taken and shows that the structure of the Predicted beam can still display deviations from the Actual beam, even at the lowest RMSE.

\begin{figure} 
\centering
\makebox[0.33\textwidth][l]{\small (a)}
{\includegraphics[width=0.33\textwidth]{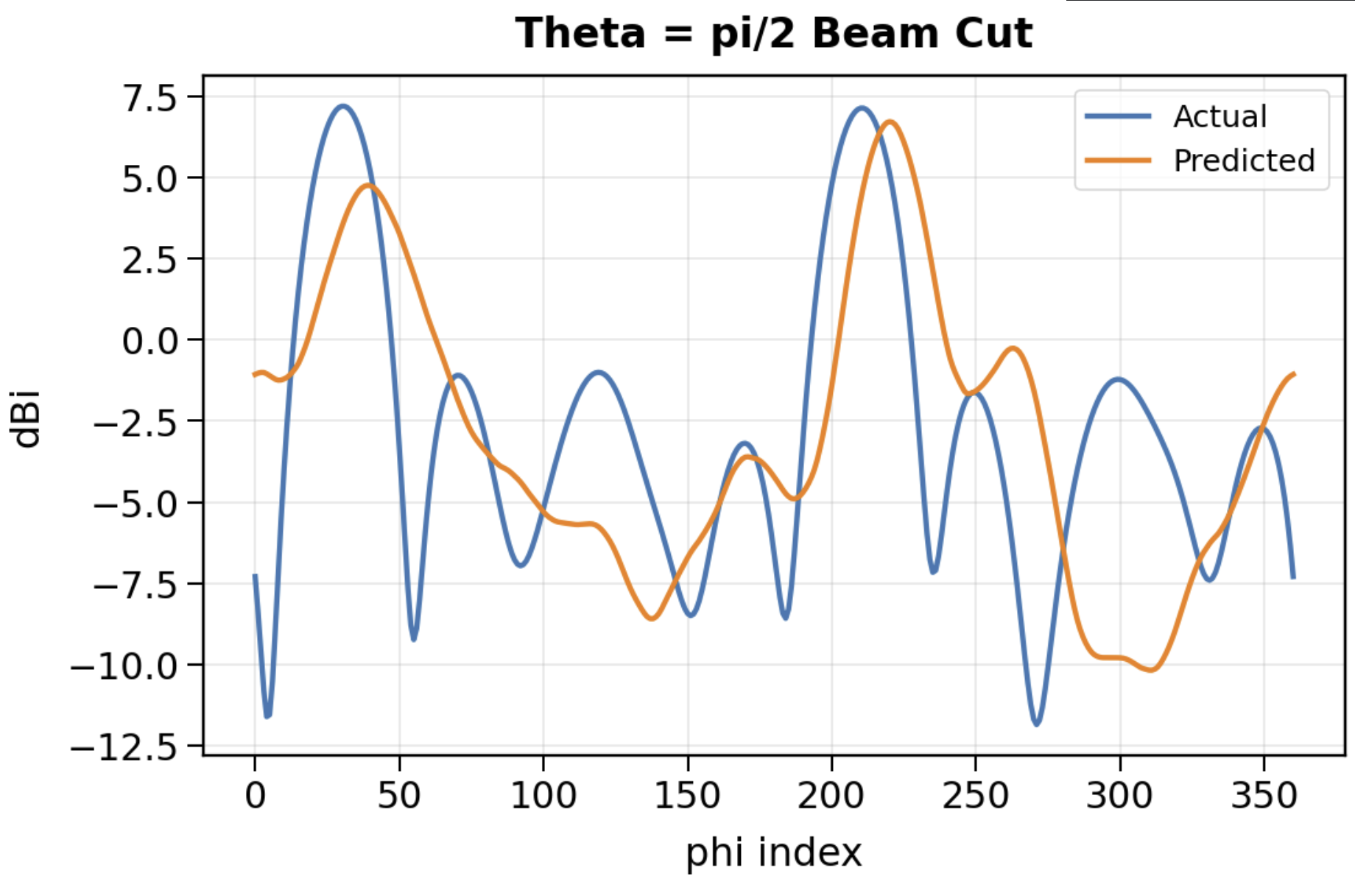}}
\makebox[0.33\textwidth][l]{\small (b)}
 {\includegraphics[width=0.33\textwidth]{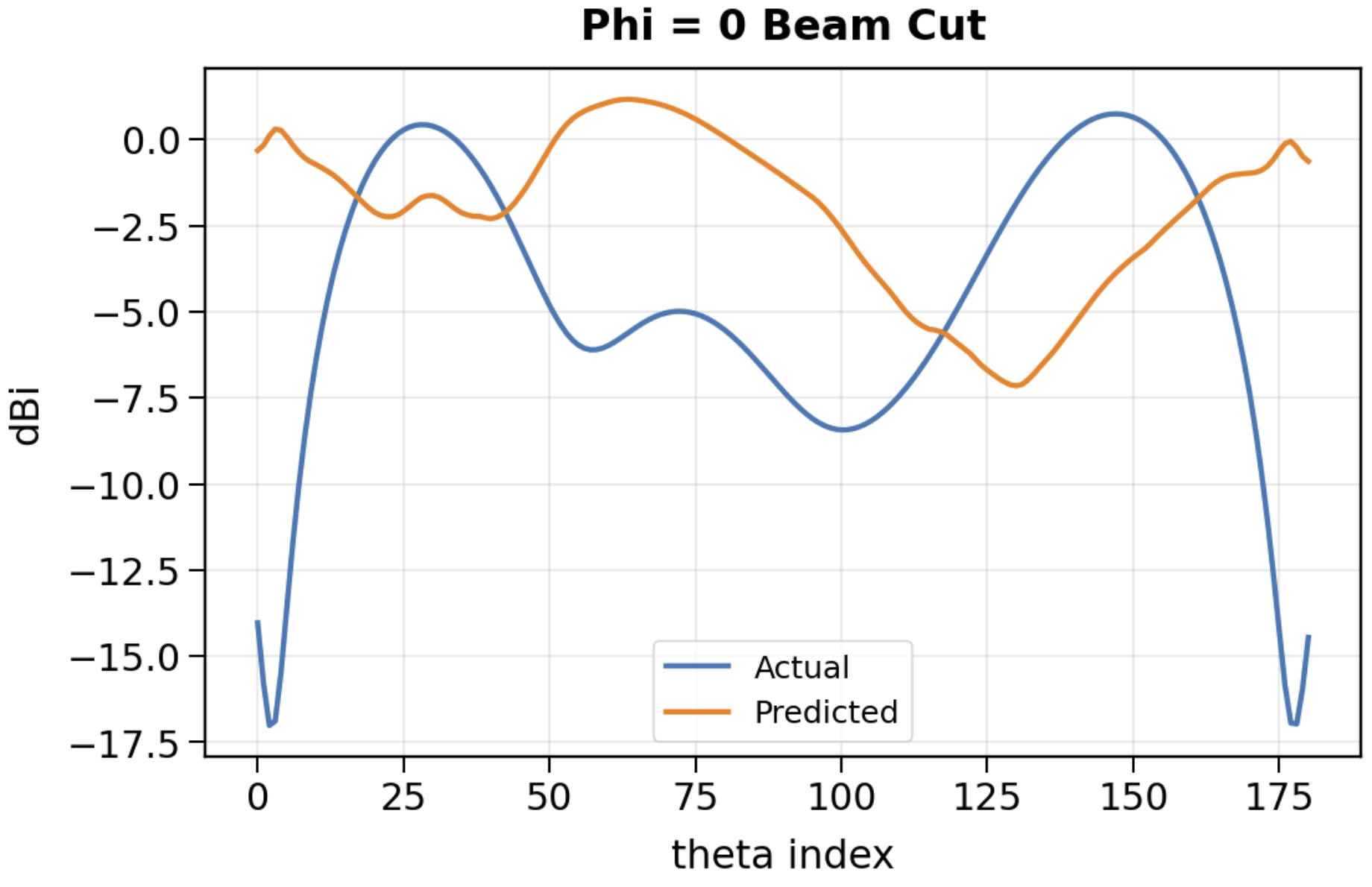}}
\makebox[0.33\textwidth][l]{\small (c)}
{\includegraphics[width=0.2\textwidth]{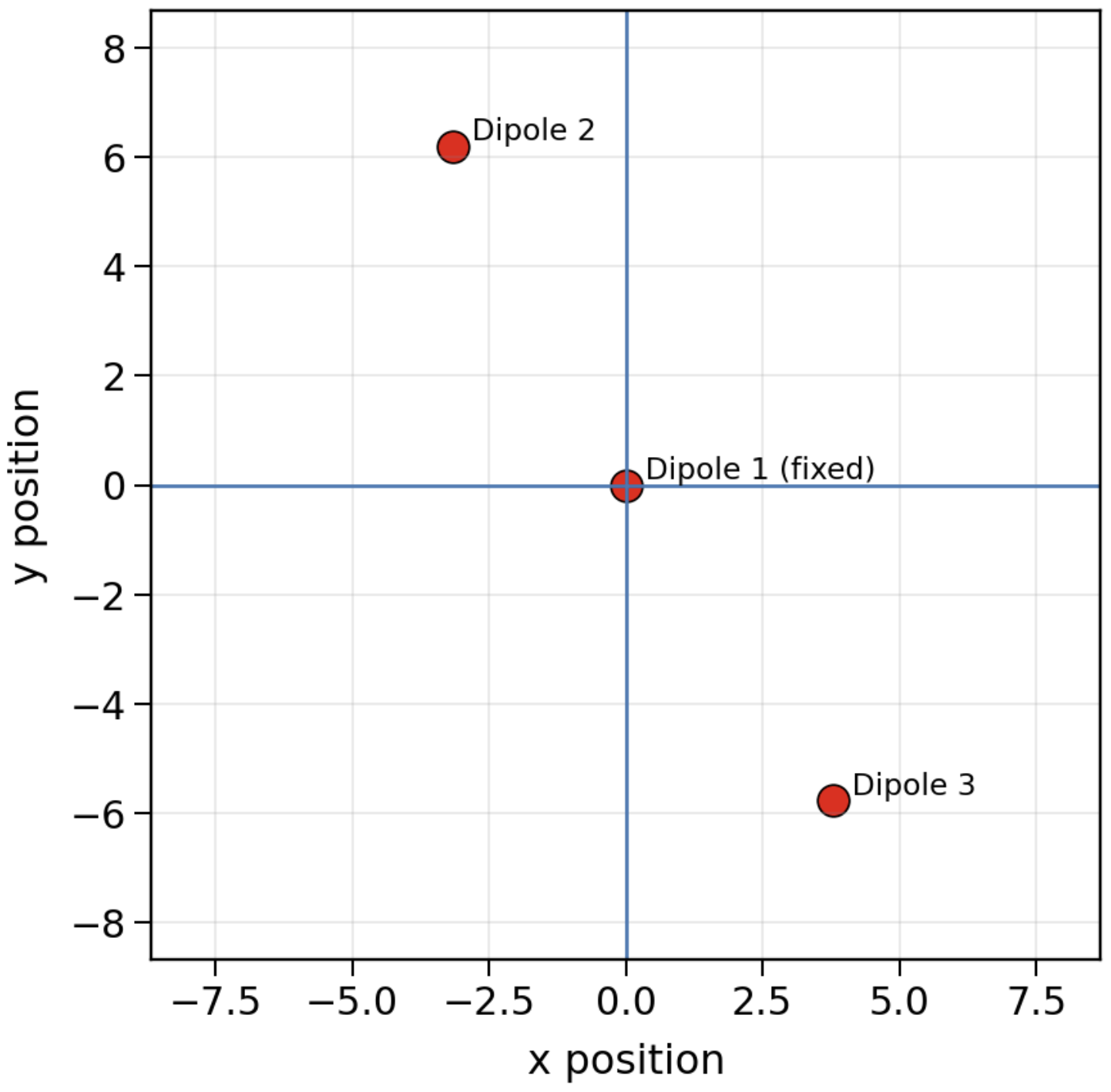}}
 {\includegraphics[width=0.48\textwidth]{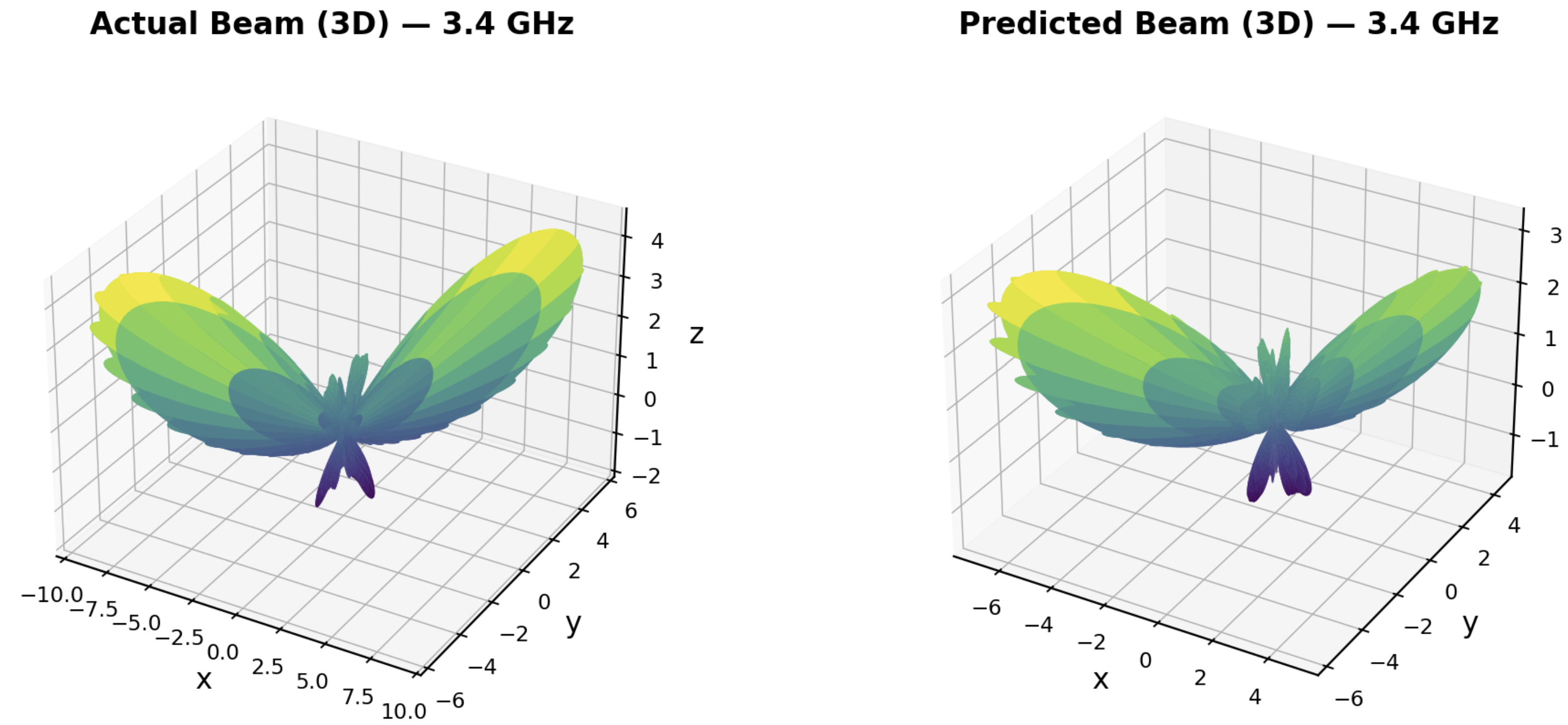}}
{\small (d)}
\caption{Array beam pattern that had the highest RMSE for case A. Panel (a) is the $\theta = \pi/2$ cut and shows the the model's difficulty in predicting the $\phi$-dependence of the pattern. Panel (b) is the $\phi = 0$ cut and shows the model's inability to predict the null points near the poles of the pattern. Panel (c) shows the antenna positions, and panel (d) depicts the general 3D structure of the Actual ({left}) and Predicted ({right}) beams. Axis units are dBi.}
\label{fig:worst_beam}
\end{figure}

Finally, Fig. \ref{fig:worst_beam} depicts the same series of plots as Fig. \ref{fig:best_MSE} for the Prediction with the highest RMSE. The Actual beam pattern varies rapidly as a function of $\phi$, which likely made this beam difficult for the DNN to predict accurately. Additionally, Fig. \ref{fig:worst_beam}(b) shows that the model struggles to predict the null points near the nulls of a beam pattern. Figure \ref{fig:worst_beam}(d) depicts the Actual vs Predicted full 3D beam, and although the general structure was still predicted relatively well, the magnitudes differ significantly.

\subsection{Errors in antenna positions}
\label{sec:incorrect}

\begin{figure} 
    \centering
    {\includegraphics[width=0.8\linewidth]{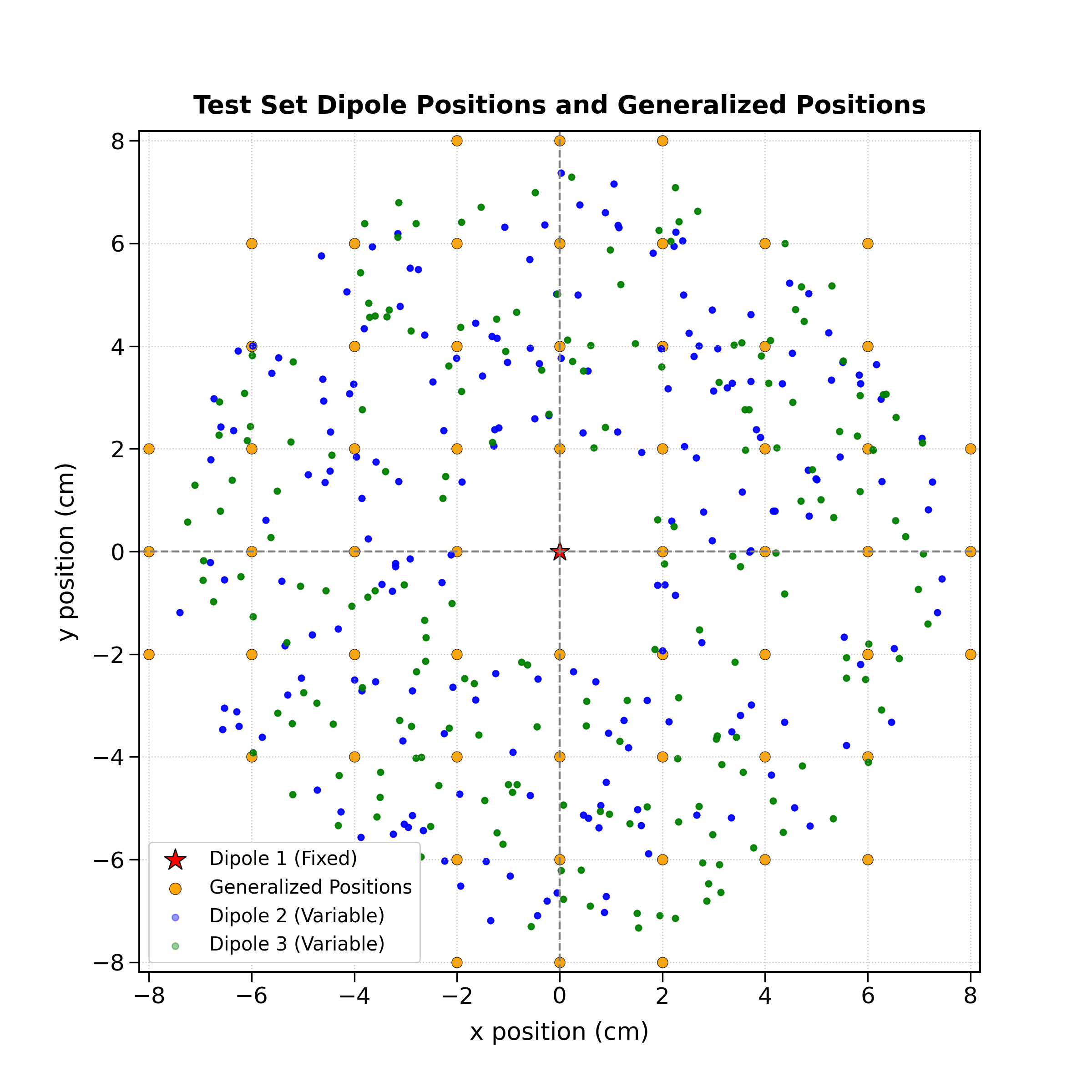}}
\caption{For case B, a display of all of the dipole positions in the test data block, along with the generalized positions that they were fit to. The orange dot closest to a dipole's true position would have that orange dot's position input to the model, which was then used to predict the beam pattern. This means we are giving the model a position value that could be incorrect by a maximum of 1~cm in both the x and y directions.}
\label{fig:dipole_locations_bad_pos}
\end{figure}

In case B, we determined how well the beams can be recovered if the S-parameters are provided accurately but the antenna positions are specified only within some range of uncertainty. For this test, the DNN training steps remained the same as \S\ref{sec:method}\ref{sec:training}, but during testing the DNN was given generalized dipole positions. Specifically, the dipole positions input to the DNN were the coordinates of the closest generalized position, shown as orange points in Fig. \ref{fig:dipole_locations_bad_pos}, to the actual antenna's position, shown as blue and green points in Fig. \ref{fig:dipole_locations_bad_pos}. The grid points are equally spaced by 2~cm in the x and y directions, meaning the maximum error in a dipole's position is 1~cm ($\sim \lambda/3$) in each direction.


Tab. \ref{tab:CasesRMSE} displays the RMSE values for the subtrain, validation, and test data sets for case B. In contrast to case A, the tested data had a higher RMSE than the validation data, likely an artifact of the uncertainty in antenna positions. However, the RMSE for the test data (2.09 dBi), was only slightly higher than for case A (Tab. \ref{tab:CasesRMSE}).


For case B, a plot of the the Actual vs. Predicted directivity for all data points  
displays prediction behavior similar to the case where the correct antenna positions were provided (Fig. \ref{fig:pred_v_act}), and is not shown.  The model's difficulty with very small values was not affected by imposing uncertainties in the antenna position inputs.

\begin{figure} 
    \centering
      {\includegraphics[width=0.48\textwidth]{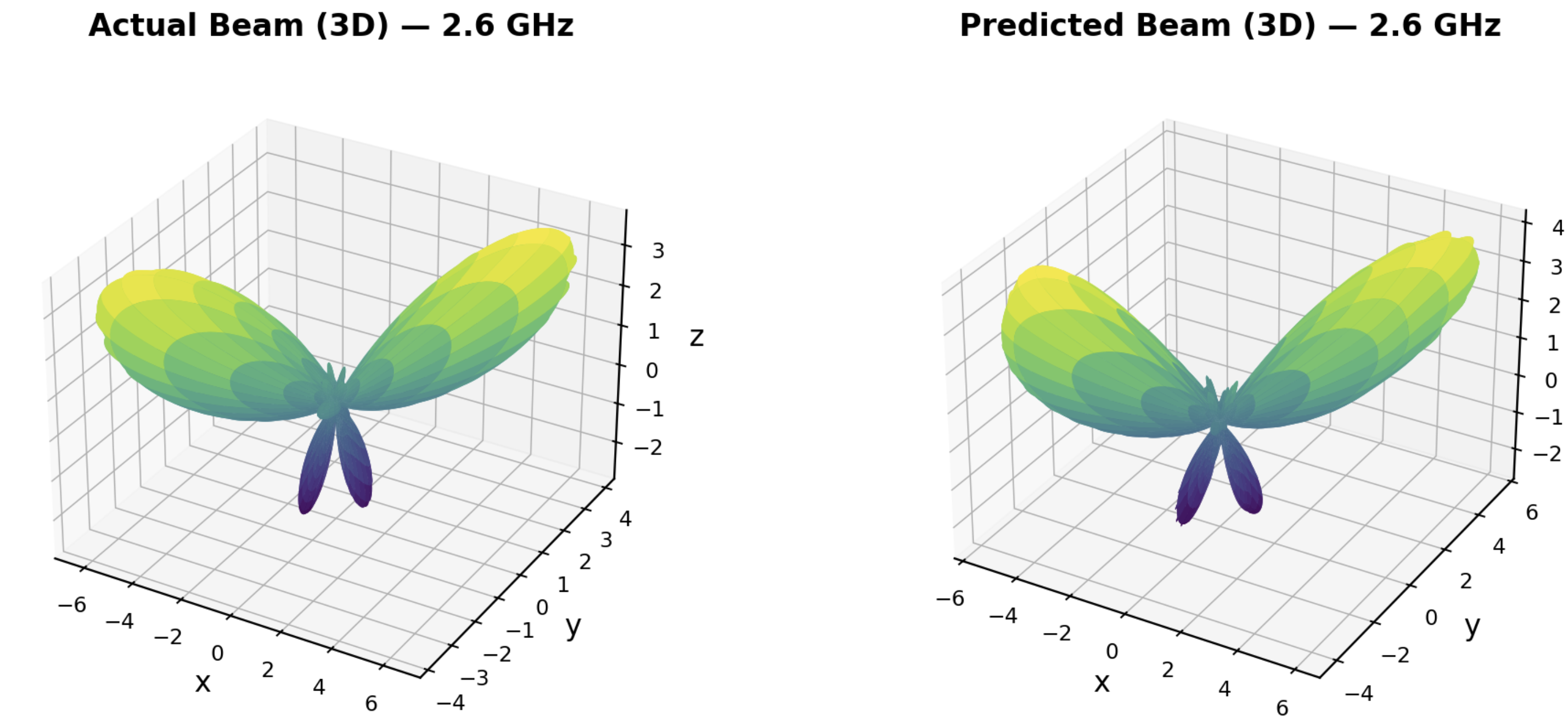}}\\
      \centering 
      {\includegraphics[width=0.48\textwidth]{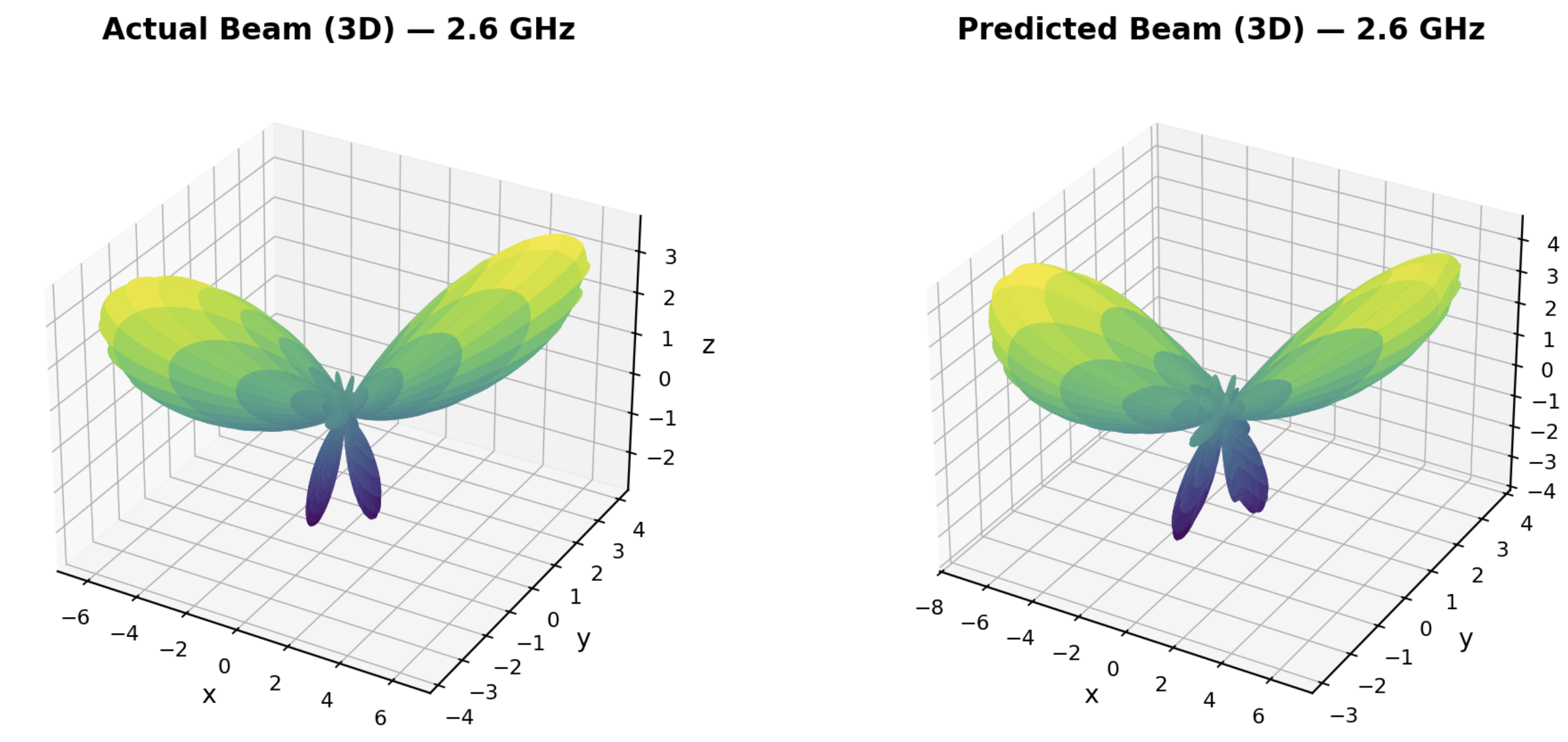}}\\
    \caption{3D visual comparisons of Actual beams (simulated, left column) and DNN-Predicted beams (right column) for case A (accurate positions, top row) and case B (inaccurate positions,  bottom row).  
      There is remarkable similarity between Actual and Predicted beams even when the DNN is given positions with errors.}
    \label{fig:antennaDiagram_compare}
\end{figure}

The 3D visual comparisons of Actual beams compared to Predicted beams for case B looks very similar to case A  for the same nominal antenna placements.  In Fig. \ref{fig:antennaDiagram_compare} (top) we show Actual vs Predicted beams using Case A adjacent to the Actual vs Predicted beams using Case B for the same antenna placement (Fig. \ref{fig:antennaDiagram_compare} (bottom).  Comparing the Actual beams on the (left) to the DNN-Predicted beams on the (right), the DNN is clearly able to predict the general structure of the beams, even with large errors in the absolute antenna position data input to the DNN.

\begin{figure} 
    \centering 
\makebox[0.33\textwidth][l]{\small (a)}
{\includegraphics[width=0.33\textwidth]{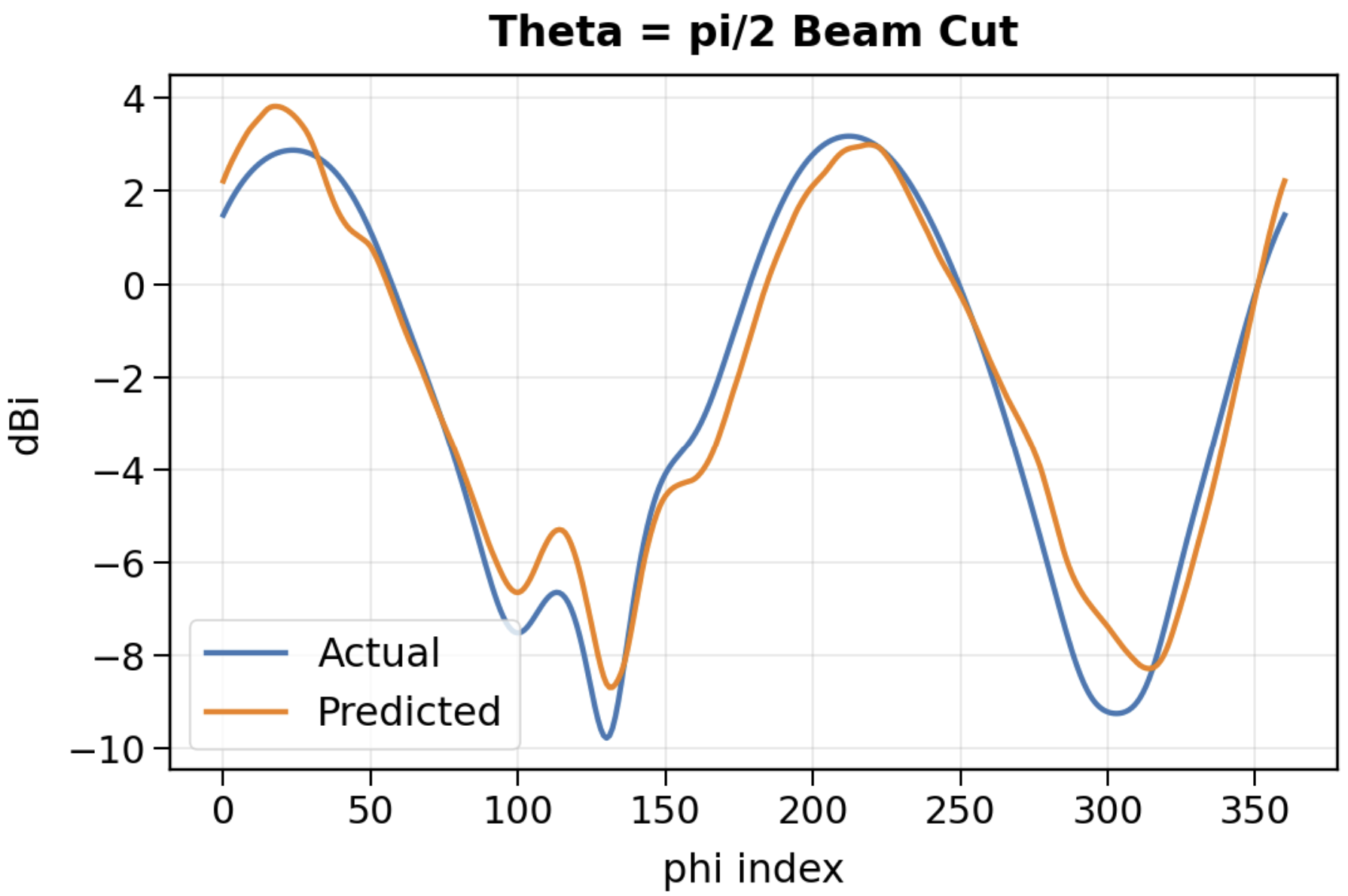}}\\
\makebox[0.33\textwidth][l]{\small (b)}
   {\includegraphics[width=0.33\textwidth]{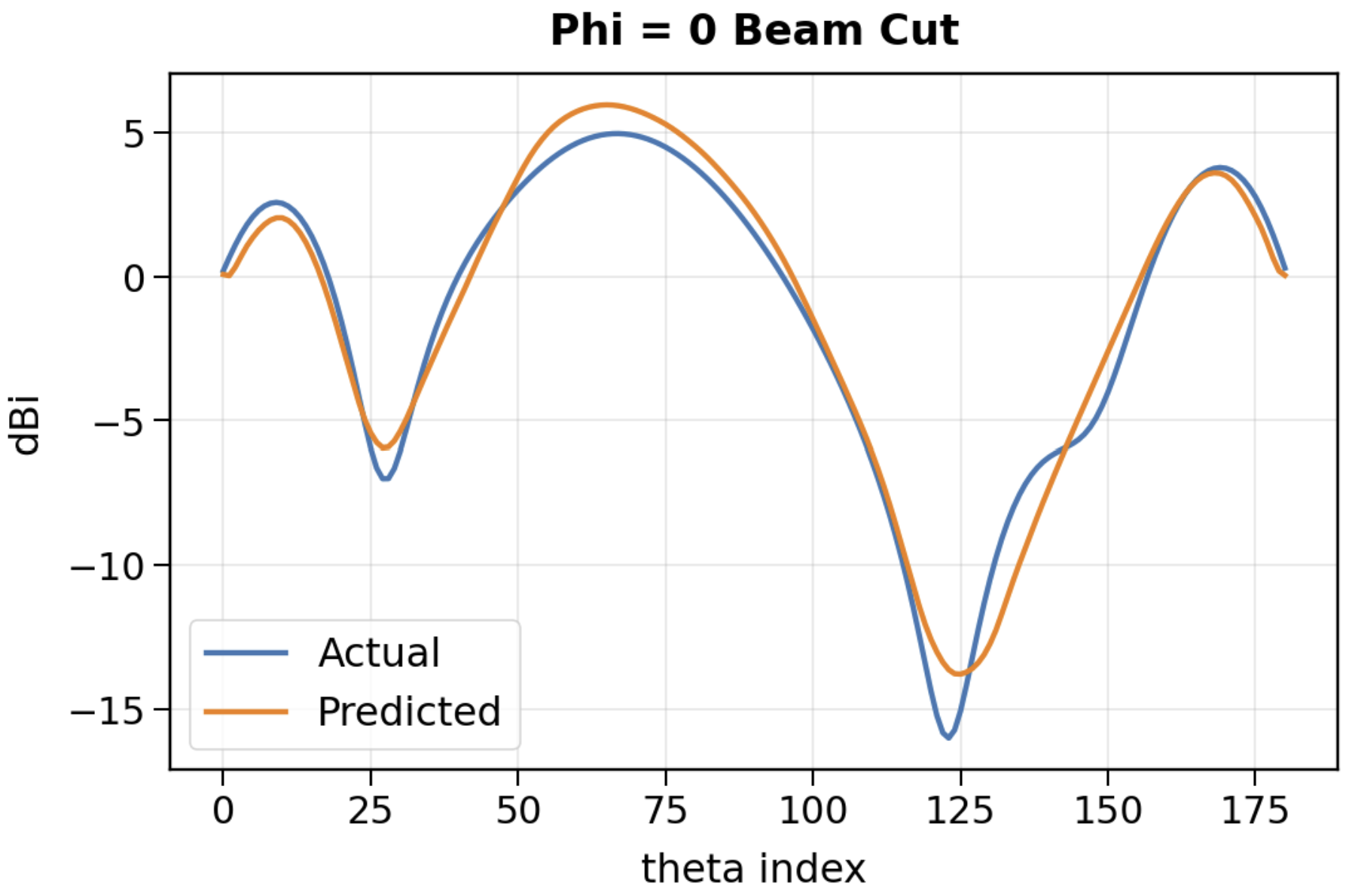}}\\
\makebox[0.33\textwidth][l]{\small (c)}
 {\includegraphics[width=0.2\textwidth]{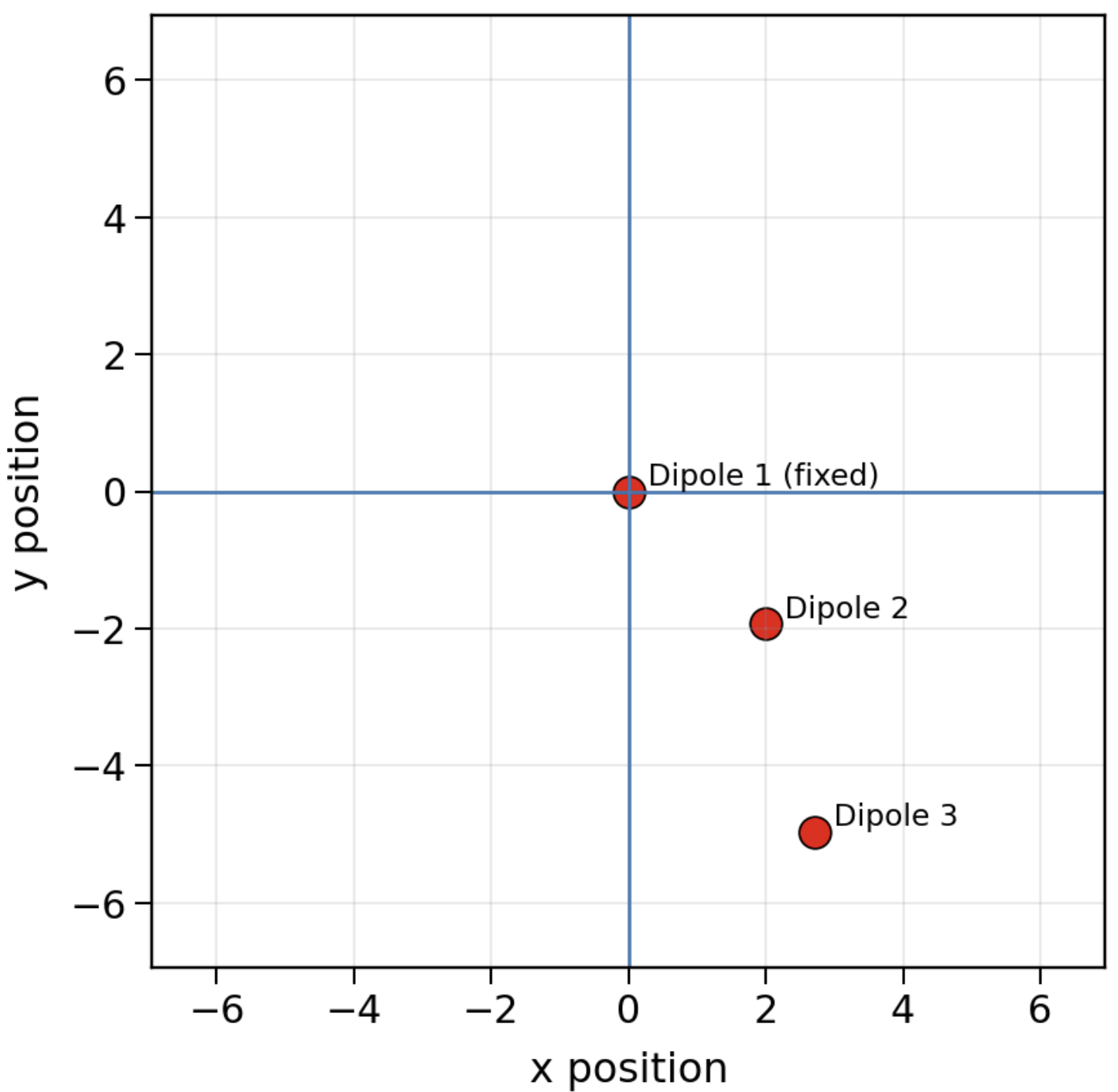}}
  {\includegraphics[width=0.48\textwidth]{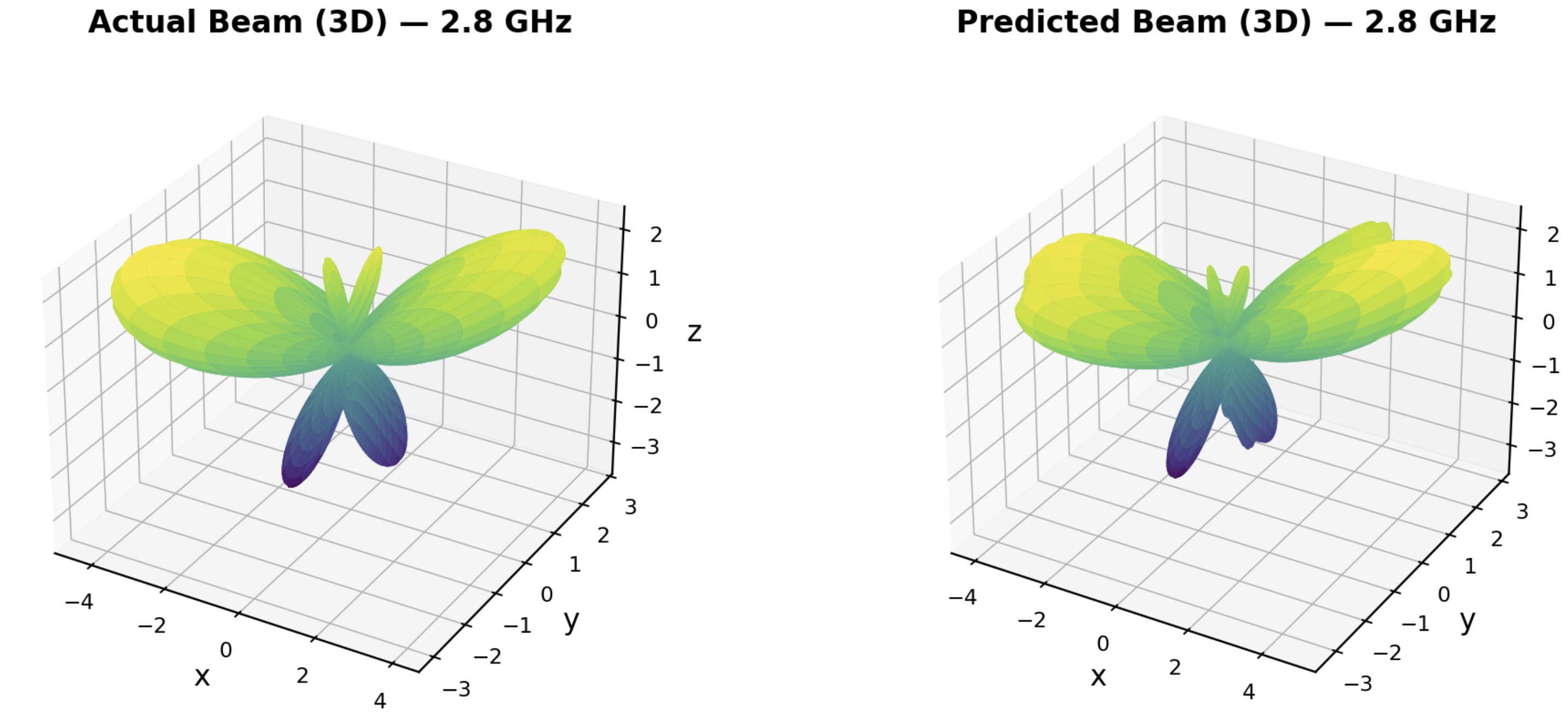}}\\
{\small (d)}
\caption{Beam patterns for one set of antenna positions that had the best agreement (lowest MSE)  between the beam pattern Predicted by the DNN and the Actual beam (simulated by Tidy3D) for Case B, where error is deliberately introduced into the antenna positions.  The beam patterns are functions of the usual 3-dimensional polar coordinates $\theta$ and $\phi$, with $\theta = 0$ oriented along the $z$-axis.  Panel (a) displays the $\theta = \pi/2$ cut through the pattern and panel (b) shows the $\phi = 0$ cut. Panel (c) shows the antenna positions and panel (d) shows  the Actual (left) and Predicted (right) 3D plots.}
\label{fig:best_beam_bad_pos}
\end{figure}

\begin{figure} 
\centering
\makebox[0.33\textwidth][l]{\small (a)}
{\includegraphics[width=0.33\textwidth]{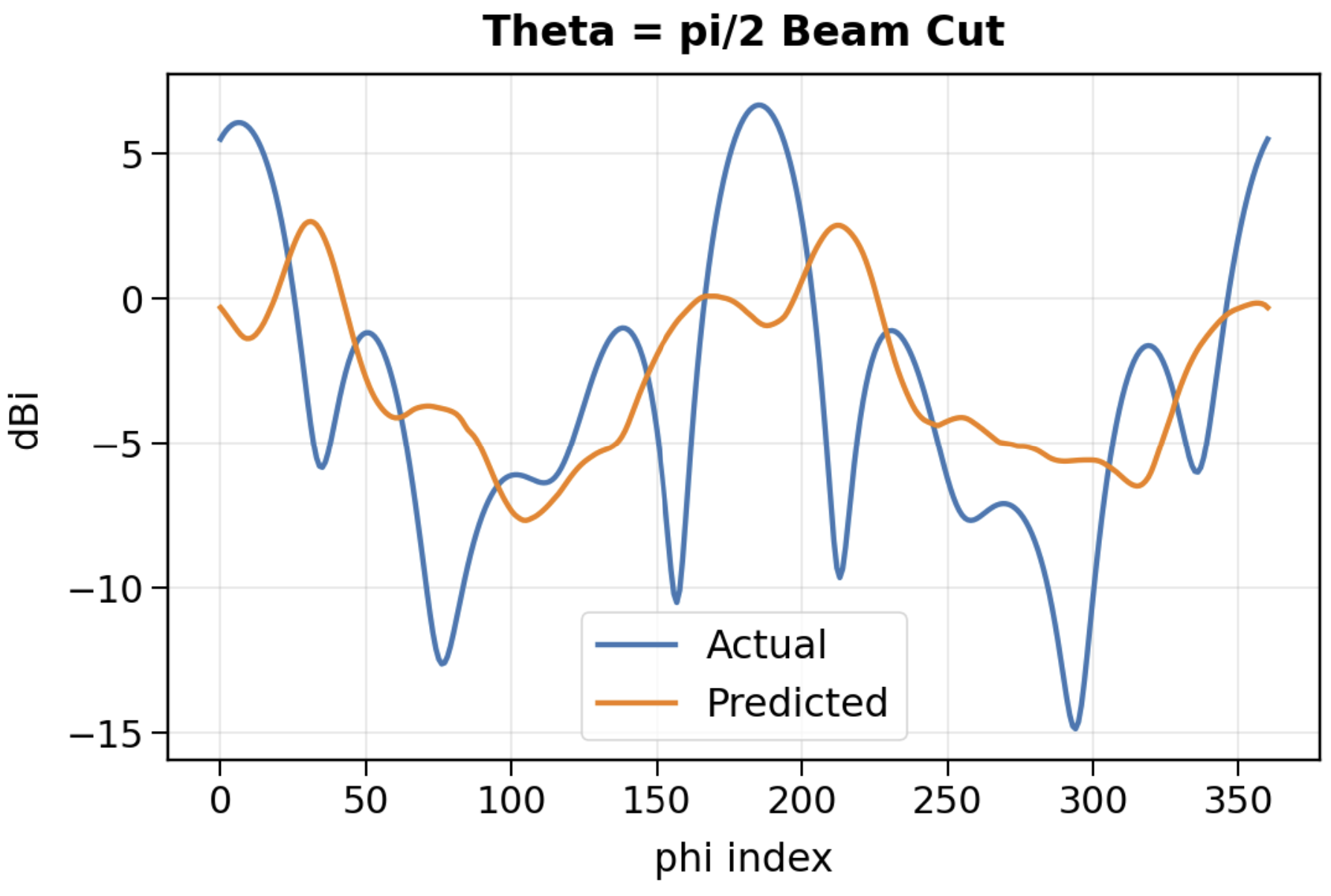}}\\
\makebox[0.33\textwidth][l]{\small (b)}
{\includegraphics[width=0.33\textwidth]{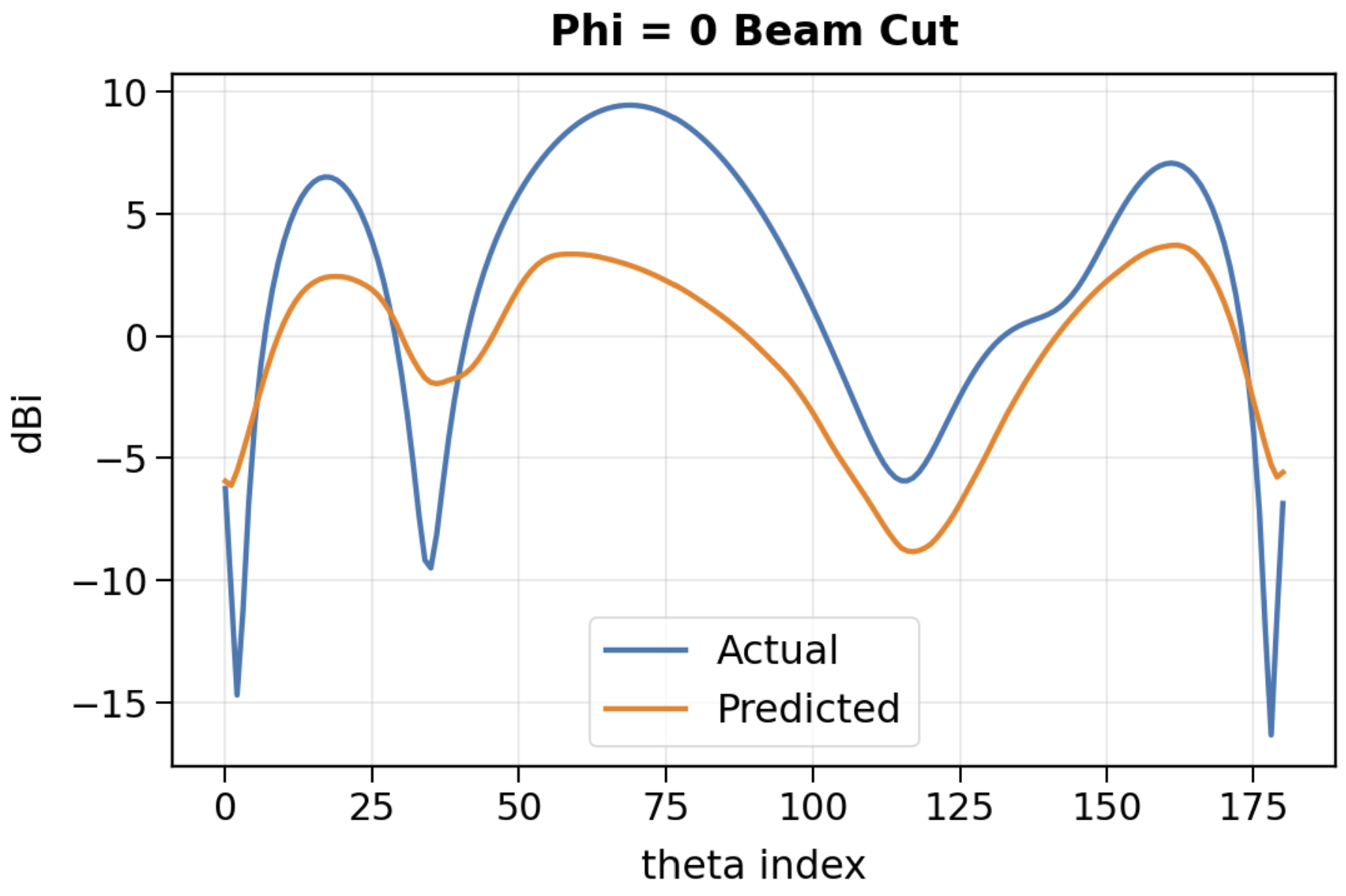}}\\
\makebox[0.33\textwidth][l]{\small (c)}
{\includegraphics[width=0.2\textwidth]{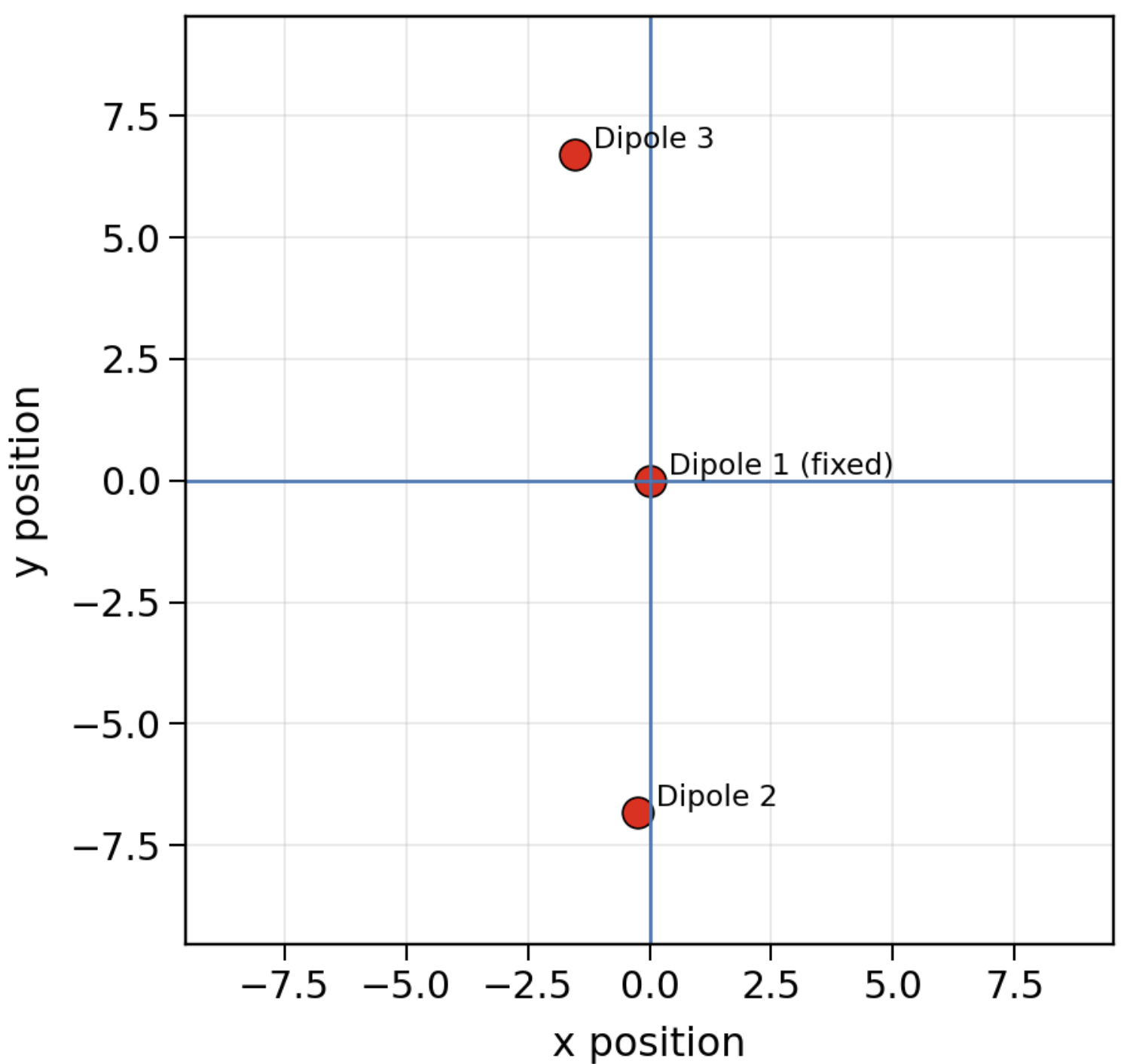}}
{\includegraphics[width=0.48\textwidth]{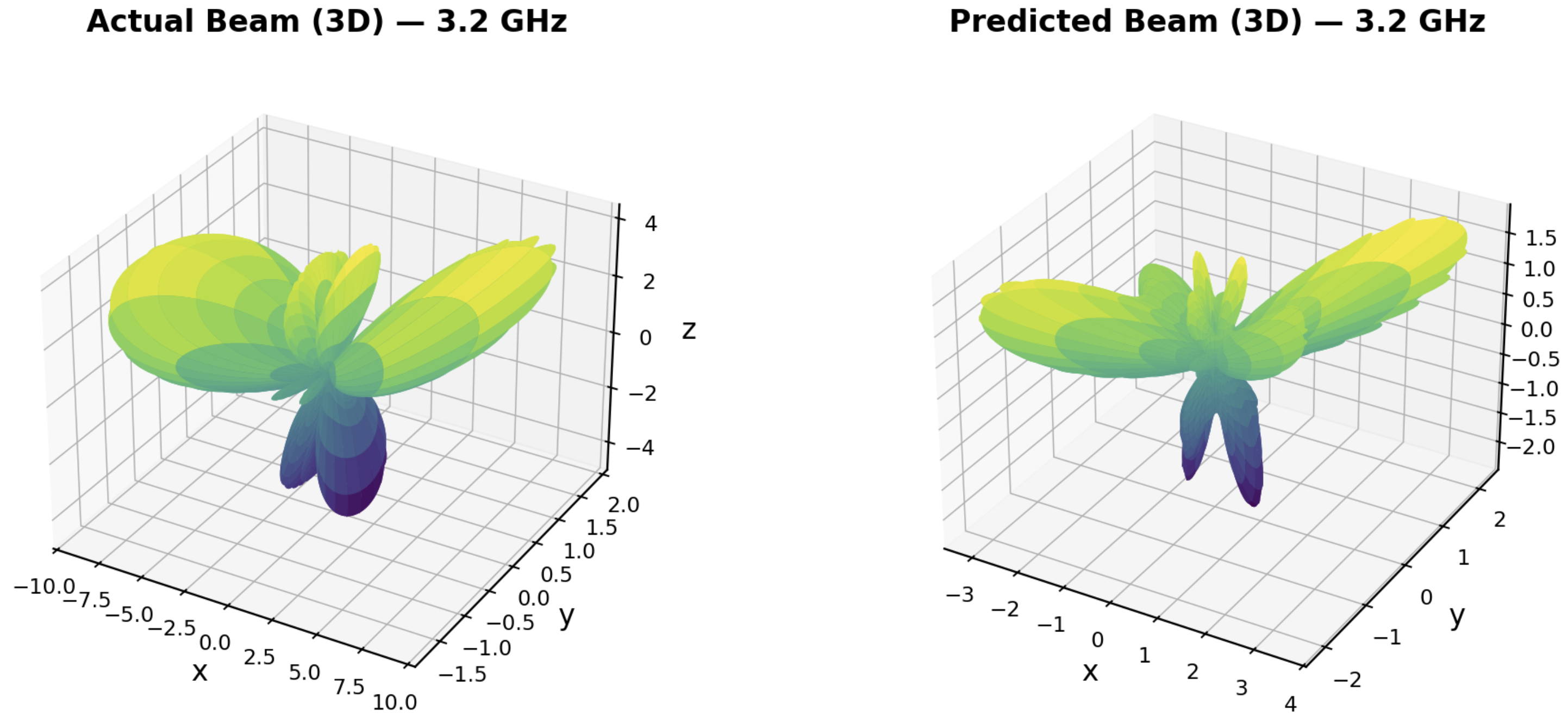}}\\
{\small (d)} 
\caption{Same as Fig. \ref{fig:best_beam_bad_pos}, but now for the set of antenna positions that gave the worst agreement between Predicted and Actual beams for case B.}
\label{fig:worst_beam_bad_pos}
\end{figure}

Lastly, the Predicted beams for case B with the lowest and highest MSE values can be seen in Figs. \ref{fig:best_beam_bad_pos} and \ref{fig:worst_beam_bad_pos}, respectively. The close agreement between the Predicted and Actual beams in Fig. \ref{fig:best_beam_bad_pos} illustrates the DNN's ability to continue to accurately predict beam patterns across a wide range of power and angles, even when given incorrect antenna positions. The beam with the highest MSE, seen in Fig. \ref{fig:worst_beam_bad_pos}, displays very similar characteristics to the beam with the highest MSE in the accurate antenna position test case. There was significant variation as a function of  $\phi$, as well as very low dBi values near the poles of the beam. Therefore, the cause of this high MSE value is likely not the uncertainty in the input antenna positions, but rather, as in case A, the structure of the DNN itself.

\subsection{Errors in S-Parameters}
In case C, we determined how well the beams can be recovered if we added an error of up to $\pm$ 5\% of each of the original S-parameter values. For this test, the DNN training steps remained the same as \S II\ref{sec:training}, but during testing, the real and imaginary components of each S-parameter given to the DNN were independently multiplied by a uniformly distributed random number between 0.95 and 1.05.

Table \ref{tab:CasesRMSE} displays the RMSE values for the training, validation, and test data sets for case C.  Similar to case A, but in contrast to case B, the tested data had a RMSE similar to the validation data.   The RMSE for the test data was 1.9 dBi, only slightly higher than case A (Tab. \ref{tab:CasesRMSE}).


For case C a plot of the the Actual vs. Predicted directivity for all data points  
is again similar to case A (Fig. \ref{fig:pred_v_act}) and case B  and is not shown. 

\begin{figure} 
    \centering
      {\includegraphics[width=0.48\textwidth]{figures/3d_beam_compare12.png}}\\
      \centering
      {\includegraphics[width=0.48\textwidth]{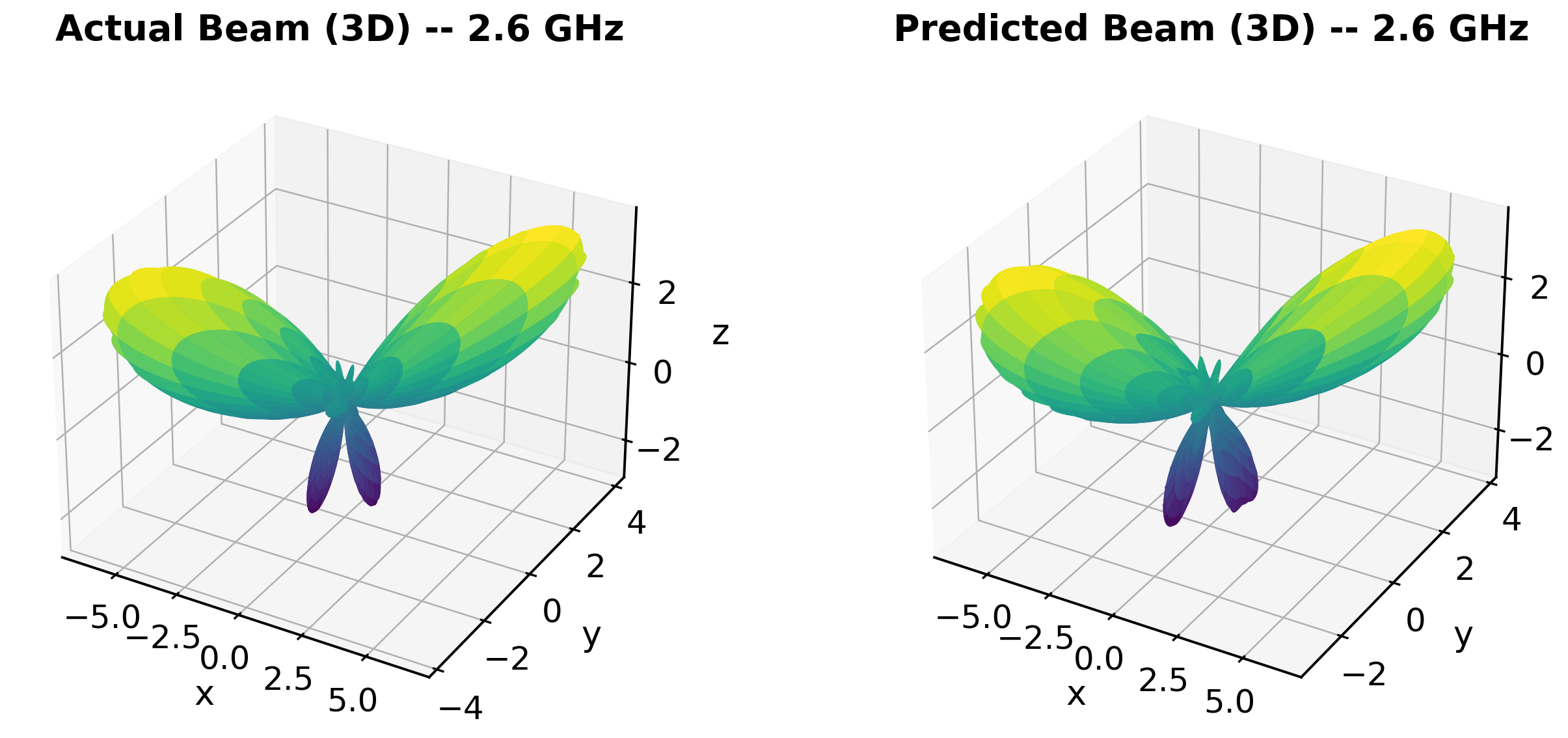}}\\
    \caption{3D visual comparisons of Actual beams (simulated, left column) and DNN-Predicted beams (right column) for case A (accurate positions and S-parameters, top row) and case C (accurate positions but 5\% uncertainty in S-parameters, bottom row).  
      Again, there is remarkable similarity between Actual and Predicted beams even when the DNN is given S-parameters with errors.}
    \label{fig:antennaDiagram_compare_BadSParams}
\end{figure}

The 3D visual comparisons of Actual beams compared to Predicted beams for case C again looks very similar to case A, where no errors were introduced.   In Fig. \ref{fig:antennaDiagram_compare_BadSParams} (top) we show Actual vs Predicted beams for the same example of case A in Fig. \ref{fig:antennaDiagram_compare} adjacent to the beam recovered for the same antenna placement, but with 5\% errors introduced into the S-parameters for that configuration (Fig. \ref{fig:antennaDiagram_compare_BadSParams}, bottom).  Comparing the Actual beams (left) to the DNN-Predicted beams (right), the DNN is clearly still able to predict the general structure of the beams, even with errors in the S-parameters input to the DNN.

\begin{figure} 
    \centering
\makebox[0.33\textwidth][l]{\small (a)}
    {\includegraphics[width=0.33\textwidth]{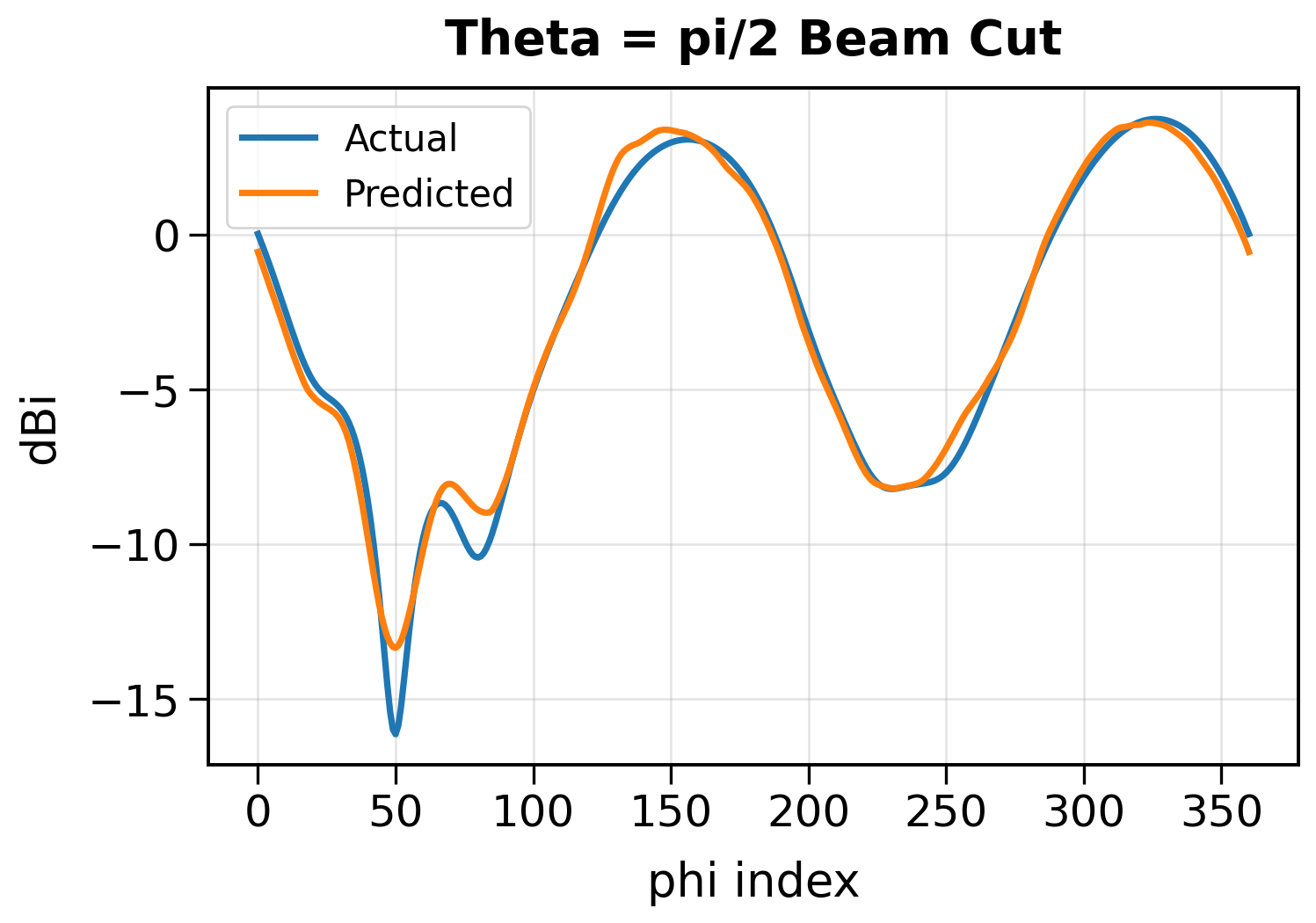}}
\makebox[0.33\textwidth][l]{\small (b)}
    {\includegraphics[width=0.33\textwidth]{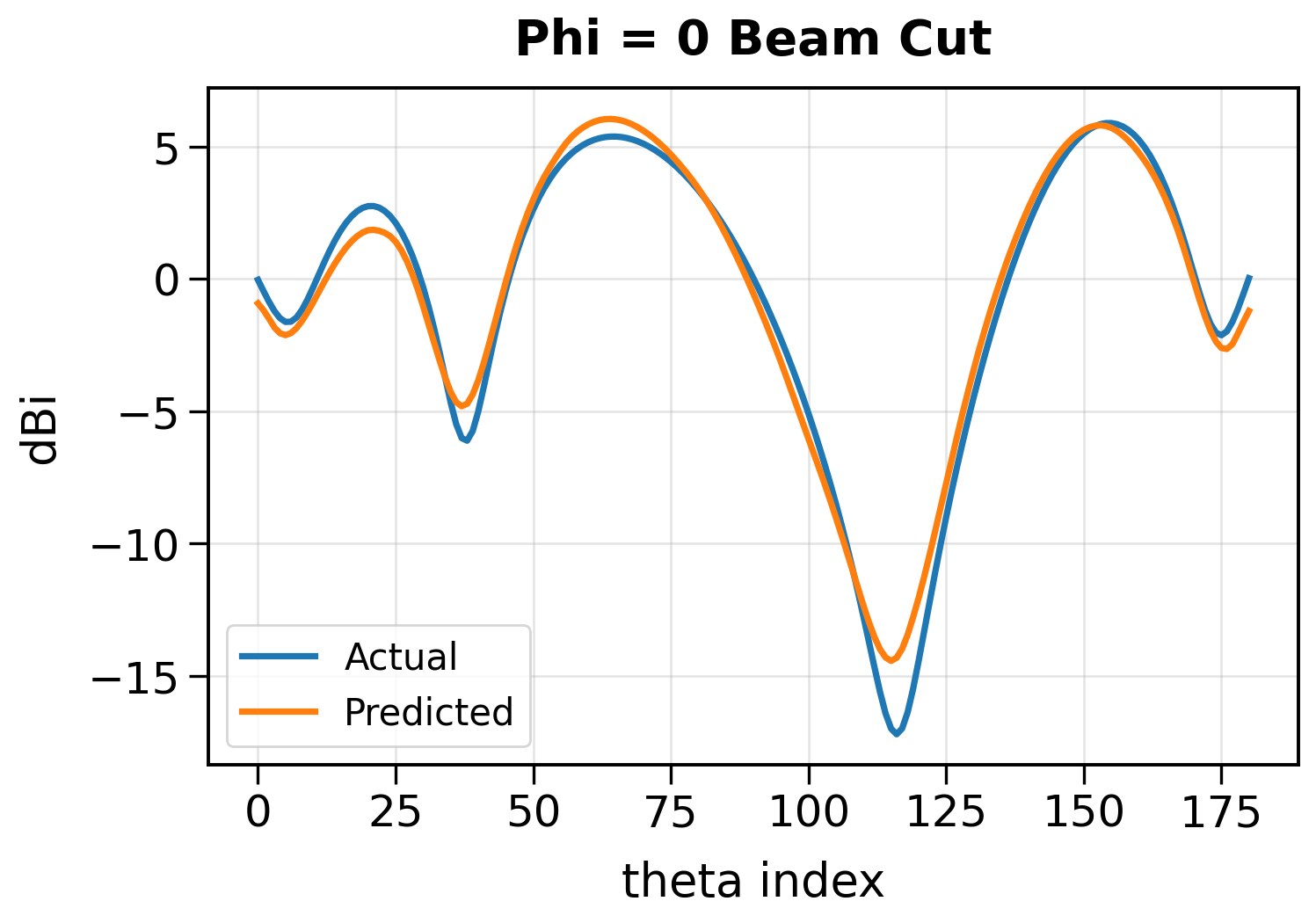}}
\makebox[0.33\textwidth][l]{\small (c)}
    {\includegraphics[width=0.2\textwidth]{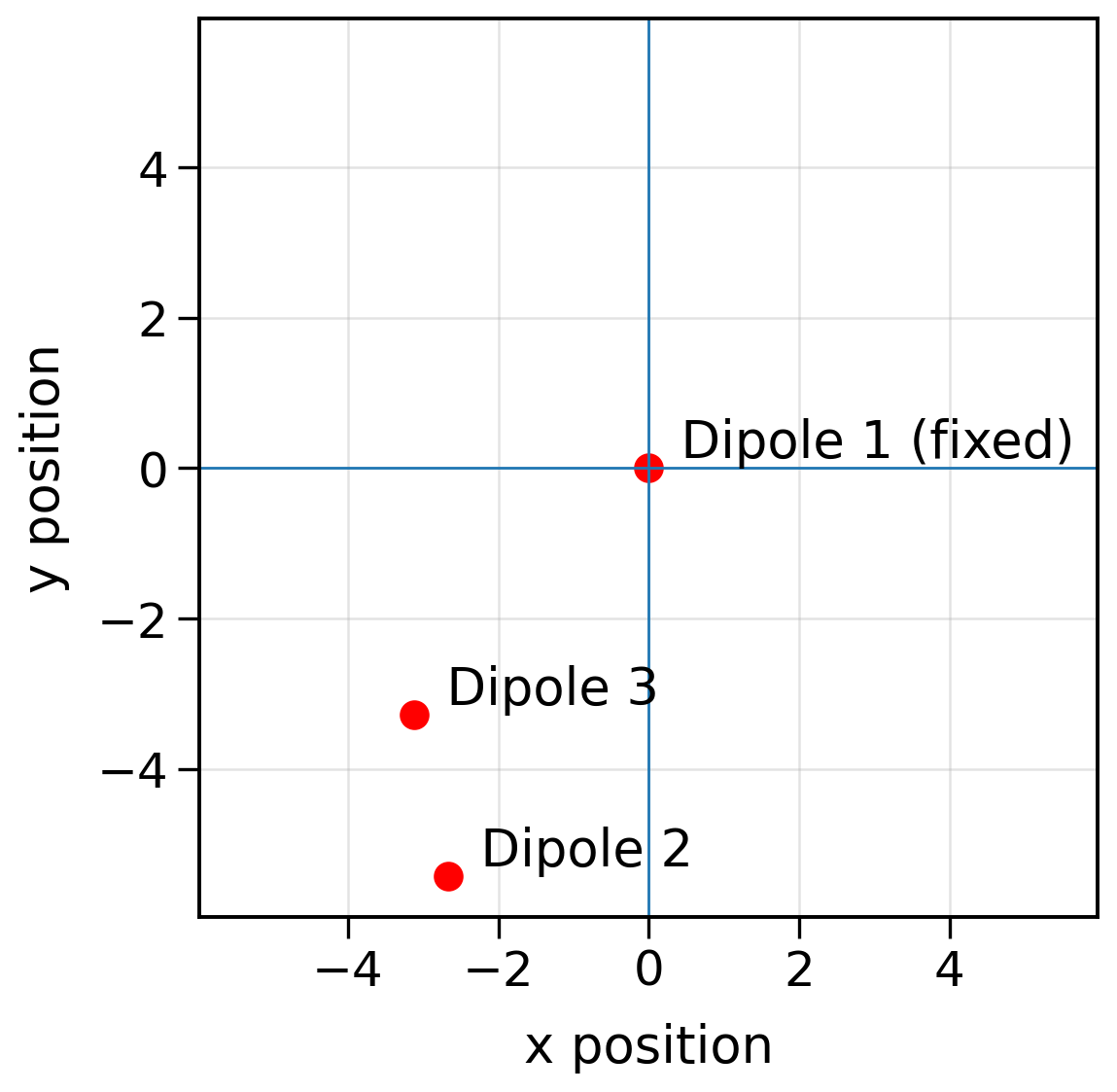}}
     {\includegraphics[width=0.48\textwidth]{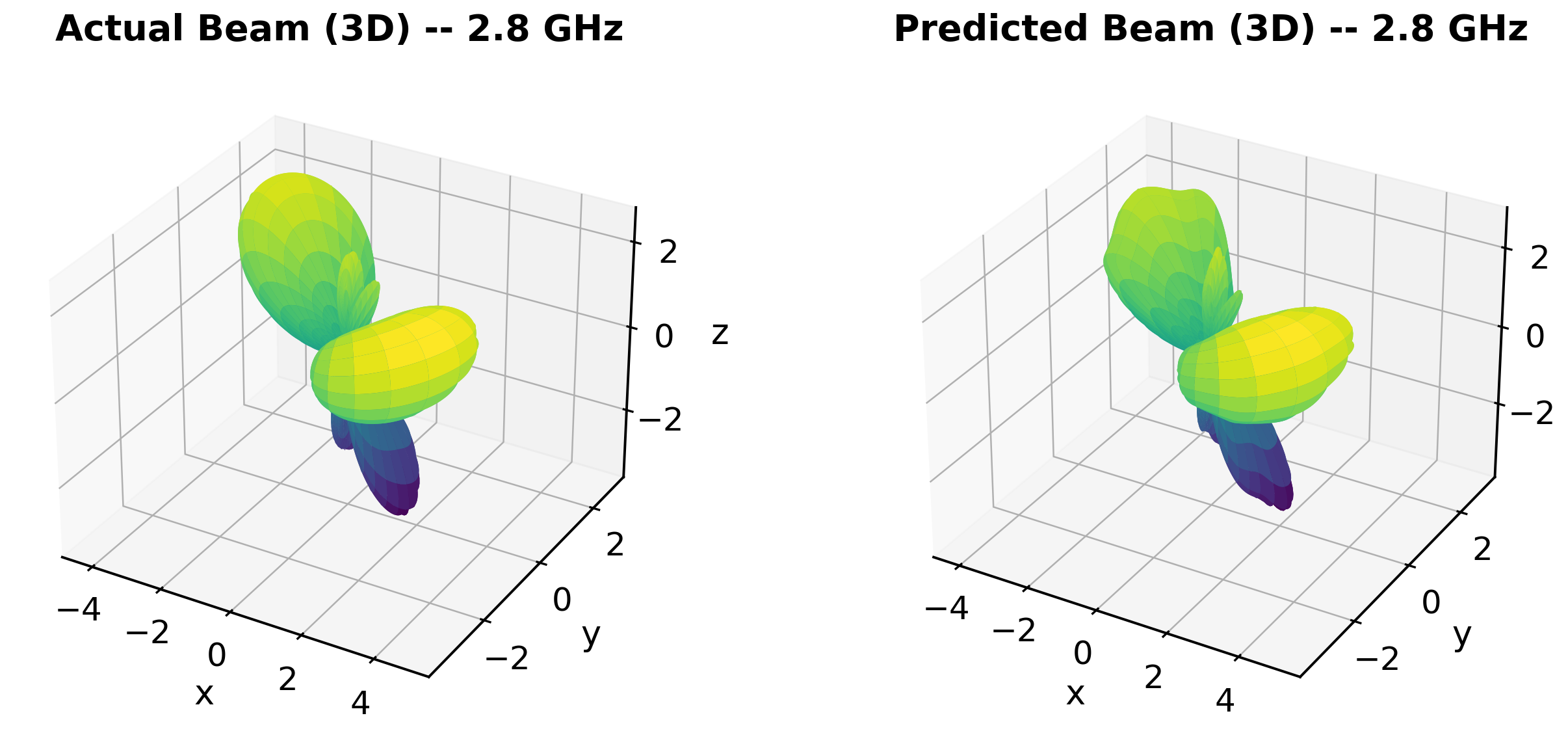}}
{\small (d)} 
\caption{Beam patterns for one particular 3-element array that had the best agreement (lowest MSE)  between the beam pattern Predicted by the DNN and the Actual beam (simulated by Tidy3D) for Case C, when 5\% errors are introduced in the S-parameters. The beam patterns are functions of the usual 3-dimensional polar coordinates $\theta$ and $\phi$, with $\theta = 0$ oriented along the $z$-axis.  Panel (a) displays the $\theta = \pi/2$ cut through the pattern and panel (b) shows the $\phi = 0$ cut. Panel (c) shows the antenna positions and in panel (d) are  the Actual (left) and Predicted (right) 3D plots.}
\label{fig:best_beam_bad_sparams}
\end{figure}

\begin{figure} 
    \centering
\makebox[0.33\textwidth][l]{\small (a)}
      {\includegraphics[width=0.33\textwidth]{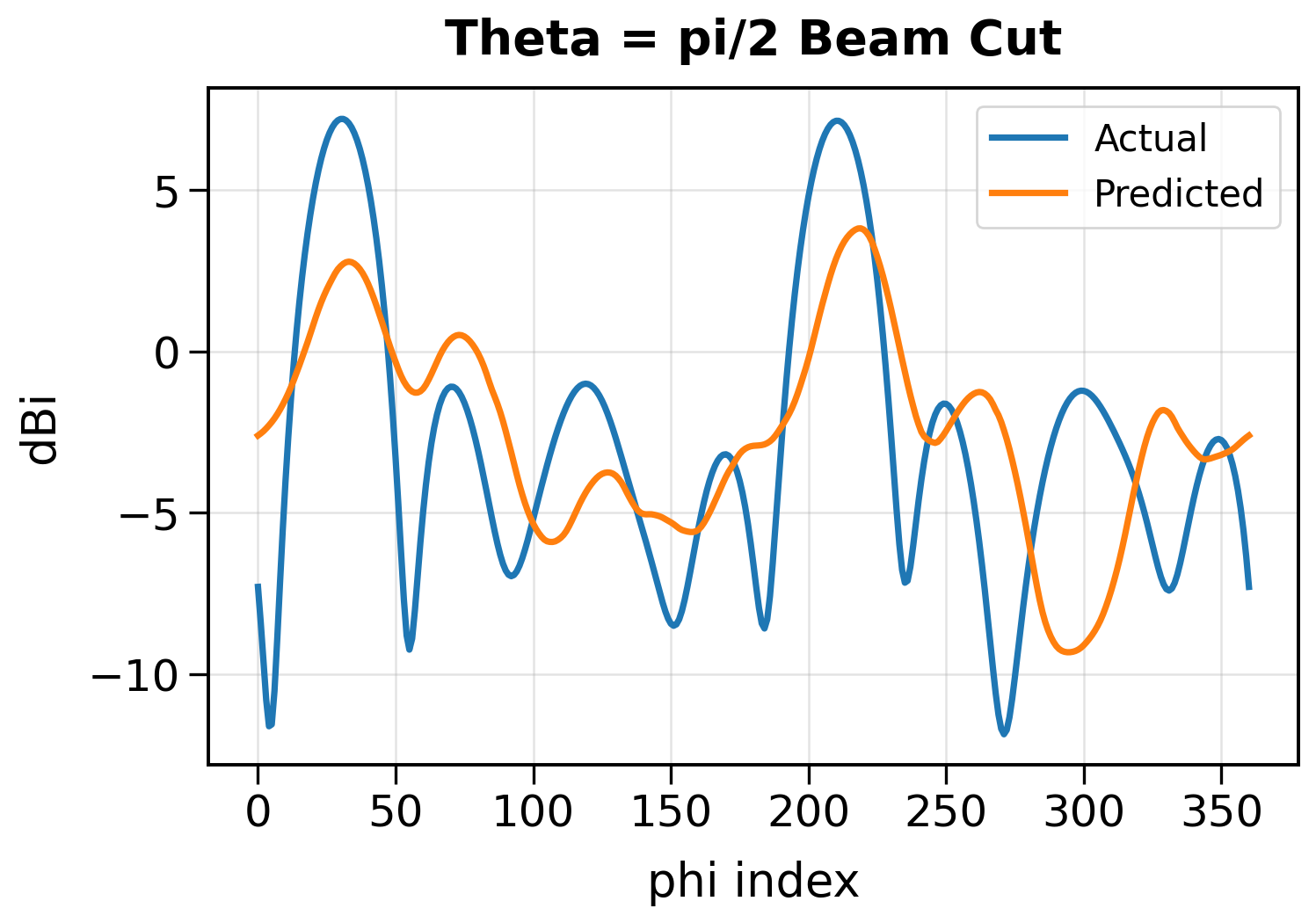}}
\makebox[0.33\textwidth][l]{\small (b)}
   {\includegraphics[width=0.33\textwidth]{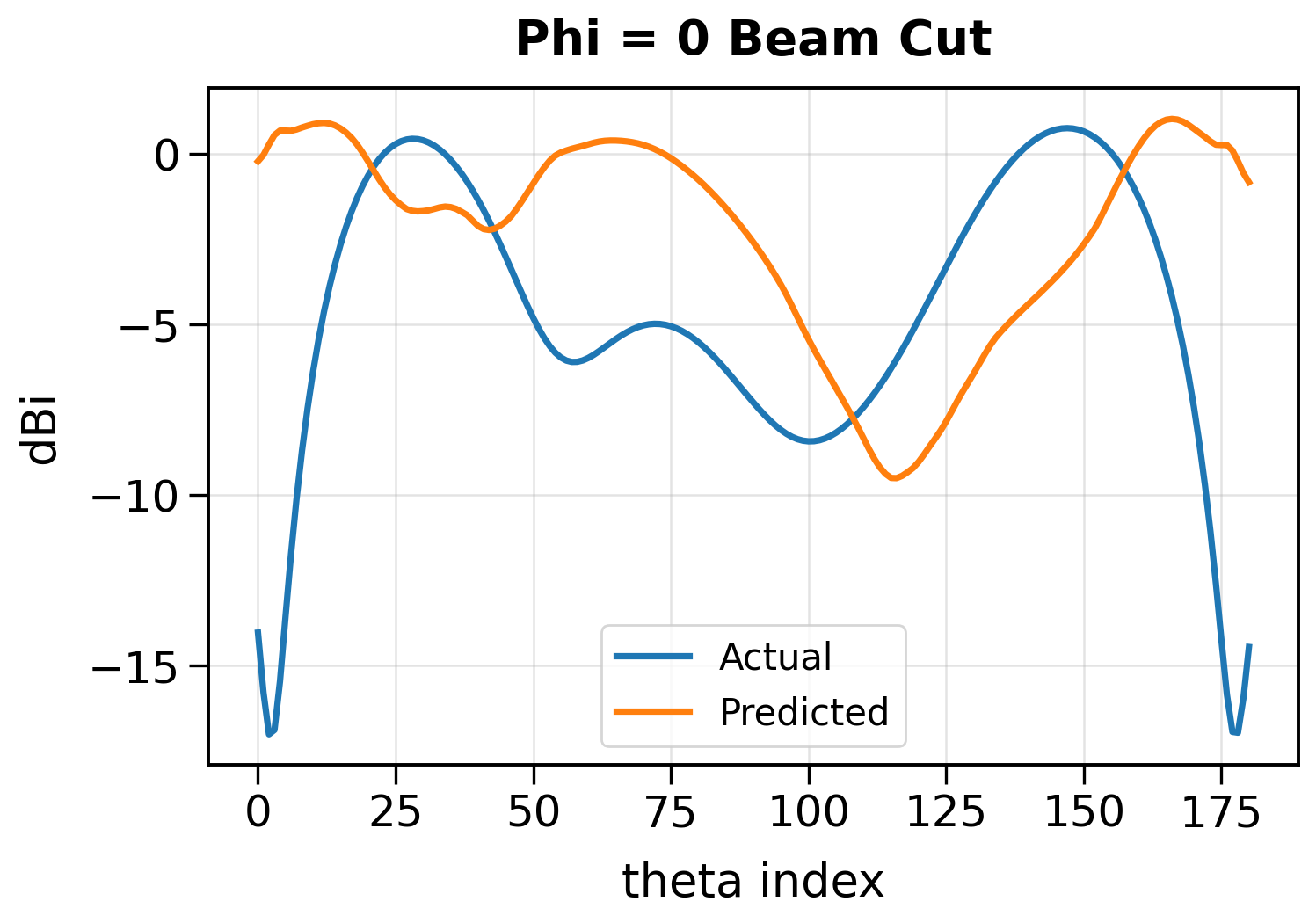}}
\makebox[0.33\textwidth][l]{\small (c)}
   {\includegraphics[width=0.2\textwidth]{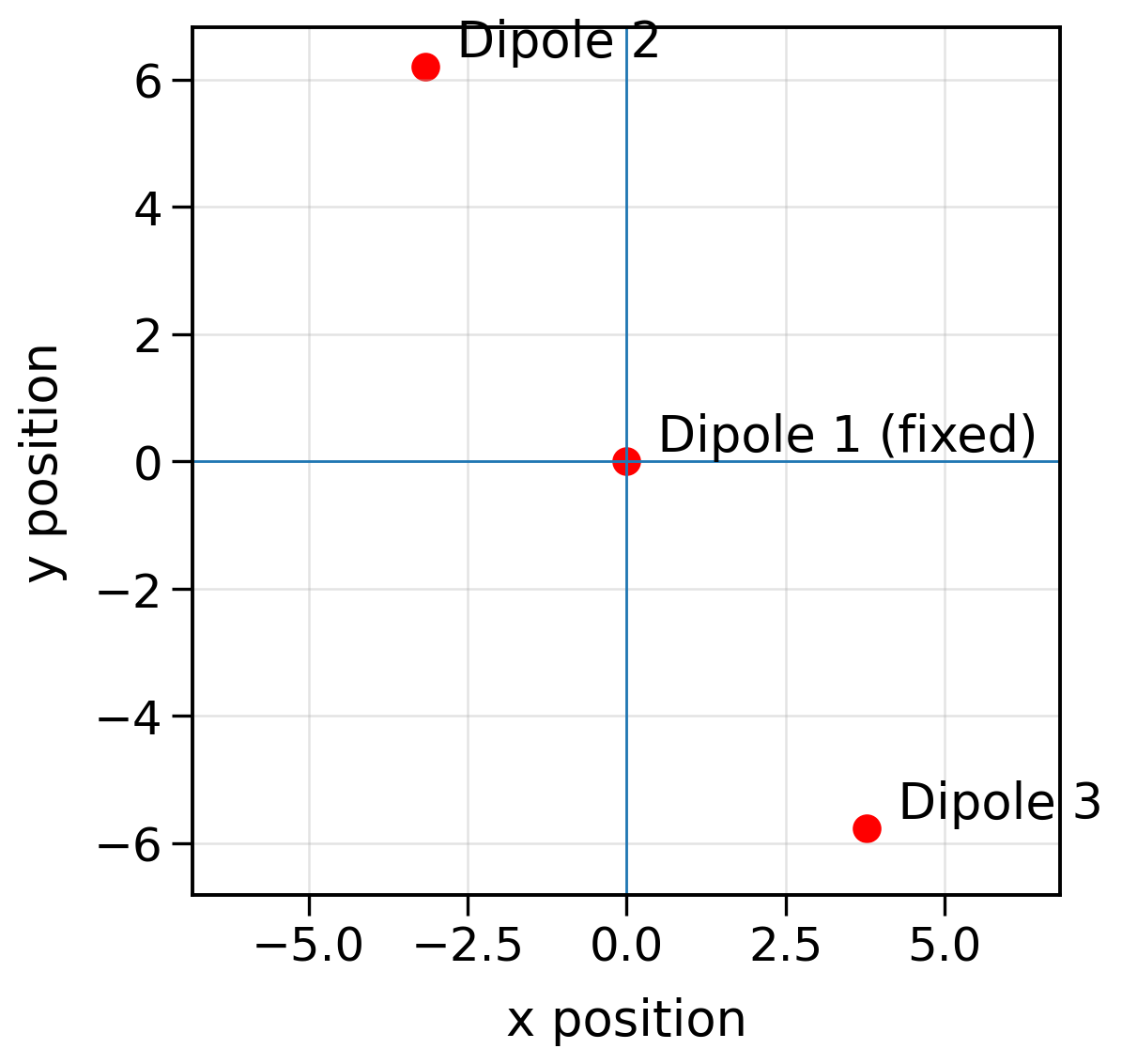}}
     {\includegraphics[width=0.48\textwidth]{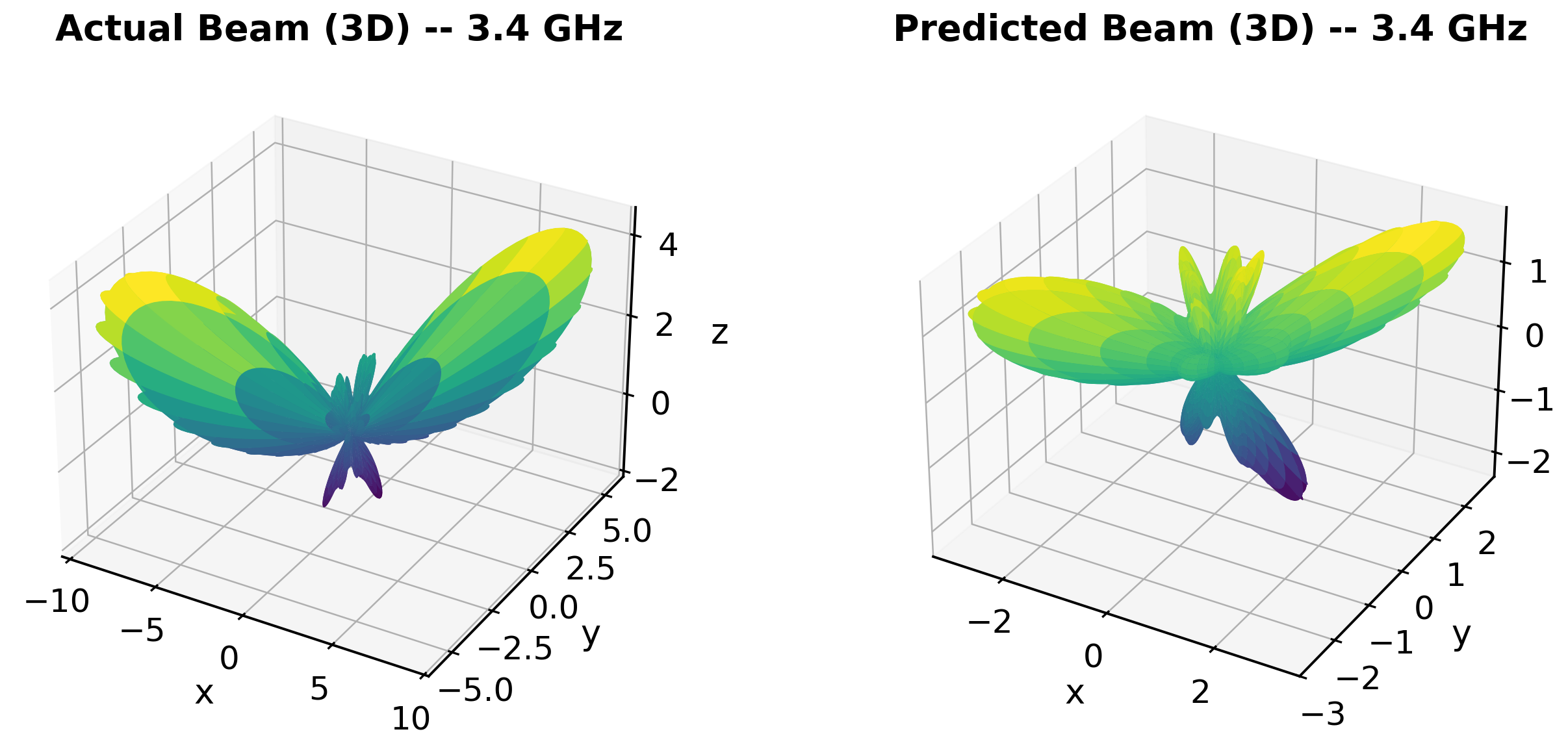}}
{\small (d)} 
\caption{Same as Fig. \ref{fig:best_beam_bad_sparams}, but now for the particular 3-element array  that gave the worst agreement (largest MSE) between Predicted and Actual beams for case C, when 5\% errors are introduced in the S-parameters.}
\label{fig:worst_beam_bad_sparams}
\end{figure}

Finally, plots of the Predicted beams for case C with the lowest and highest MSE values between the Actual and Predicted beams can be seen in Figs. \ref{fig:best_beam_bad_sparams} and \ref{fig:worst_beam_bad_sparams}, respectively.  The example of best agreement is qualitatively very similar to those for case A (Fig. \ref{fig:best_MSE}) and B (Fig. \ref{fig:best_beam_bad_pos}).   Similarly, the examples of worst agreement are qualitatively similar to case A and B (Fig. \ref{fig:worst_beam} and \ref{fig:worst_beam_bad_pos}, respectively).

To further evaluate the sensitivity of the DNN to inaccuracies in the S-parameters, additional tests were performed with perturbations of up to $\pm1$\%, $\pm 10$\%, and $\pm50$\% applied to the real and imaginary components of the S-parameters. As expected, the prediction error increased with the magnitude of the introduced errors, however the model demonstrated some robustness to these inaccuracies. The prediction error remained at 2.324 dBi RMSE even when perturbations of up to $\pm50$\% were introduced. These results suggest that the DNN can tolerate substantial relative errors in the input S-parameters, although prediction accuracy decreases as these errors increase.

\section{DISCUSSION}
\label{sec:discuss}

The previous section shows that the beam pattern of a simple 3-dipole array can be accurately predicted using ML given the antennas' positions, frequency, and S-parameters, even when the positions or S-parameters are imperfectly known.  The RMSE of the Predicted beams compared to the Actual beams in the test dataset 
was 1.89 dBi for the case where accurate antenna positions and S-parameters were provided. The RMSE degraded only slightly to 2.1 dBi when inaccurate positions ($\pm \lambda/3$) were provided and 1.9 dBi when inaccurate S-parameters ($\pm 5\% $) were provided. 
For this simple model, PCA was effective at reducing the number of parameters required to define the array beams, vastly simplifying the task of training the DNN. The Actual (simulated) beams, sampled at one-degree angular resolution over $4\pi$~sr, initially consisted of 65,341 directivity values. PCA reduced this representation to 70 coefficients, comparable to the 23 input parameters used to characterize the antenna array. The DNN learns a mapping from these 23 input parameters to the 70 PCA coefficients, which are subsequently transformed through inverse PCA to reconstruct the full 65,341-point beam patterns, spanning a dynamic range from -45 dBi to +12 dBi. 

The wide dynamic range required to describe the array patterns has posed some challenges for beam prediction.  The simulated patterns have a large, continuous range of values (-45 to +12 dBi). The DNN particularly struggled to predict values below -20 dBi (see Fig. \ref{fig:pred_v_act}), despite a significant fraction of Actual values being below -20 dBi. In general, the DNN produced Predicted directivities that were biased larger than the Actual values. This may be, in part, due to the lower values being more strongly associated with nulls in the beam pattern. Nulls occupy small fractions of the beam and may exhibit less variance across different array configurations, potentially causing PCA to neglect modes that are essential for their accurate prediction.

We tried several alternate approaches to modeling the DNN to reduce the MSE between between Predicted and Actual beams. 
One approach was a DNN which predicted the dipole excitations, including phase and amplitude, rather than the full radiation pattern. From these inputs, we analytically calculated the radiation patterns, avoiding direct beam prediction and thus the challenge of predicting 65000 output values. However, this network learned that shrinking excitation magnitudes minimized loss and therefore found an incorrect trivial solution. 

In another approach, we created a physics-informed DNN, where additional input values, such as pairwise distances, were included with the hope the model would start learning the effects of these physical properties more rapidly. This approach resulted in 13 additional input values. Unfortunately, the MSE values from this model were largely unchanged, indicating that the model had already learned these features from the original 23 inputs. 

A third alternative model was a Graph Neural Network (GNN), a network designed to process data as nodes and edges. Given that our dipole system is inherently relational, it seemed reasonable that a GNN with edge-conditioned convolutions (NNConv) would be a powerful way to predict resulting radiation patterns. In the implementation of this model, however, we were unable to reduce the RMSE values below 1.95 dBi, even after finetuning its hyperparameters with a grid search. A review of the literature revealed this was likely because our initial ``graph" consisted of exactly three dipole antennas, rather than a variable number of nodes whose interaction structure is unknown \cite{Ponzi2025}. Our original DNN model therefore outperformed these other attempted models.

In terms of computational efficiency, generating the training dataset required approximately 50 hours of full-wave FDTD simulations, with each antenna configuration taking about 3 minutes. Once trained, however, the DNN serves as a surrogate model that predicts far-field patterns from antenna positions and measured S-parameters. 

As a comparison, for an antenna array with uncertain positions, a conventional beam-prediction approach would involve a few tens of iterations of full-wave EM simulations to optimize the assumed geometry, by minimizing the discrepancy between the corresponding simulated S-parameters and measured (target) S-parameters, assuming a typical gradient-based optimization\cite{Koziel_2021}.   
In contrast, although the DNN's training dataset required 1,000 EM simulations, this upfront cost can be amortized over repeated beam-pattern predictions, potentially equivalent to 10-100 predictions beyond which the proposed approach could offer substantial computational savings, neglecting the DNN training overhead.

Although the predictions of the DNN model match the simulations well in most cases, there are a number of areas for future study, beyond the scope of this paper, for designing and understanding antenna arrays used in actual applications.
Most antenna arrays are much more complex than the three simple dipole antennas studied here. Antennas often have nontrivial shapes, environmental effects, and thousands of antennas. These arrays will require more parameters to characterize the beam patterns. Characterizing the scaling of the number of required parameters with array complexity is essential to evaluating the viability of this approach, as is developing a scaling relation between the number of antennas and the compute requirements for both the full-wave simulations and for DNN training. A logical next step would be to extend the demonstration described here to simple dipole arrays with more than three antennas.  The next level of complexity would be to evaluate the computational costs of including adding additional parameters to the training set, such as orientation of the antennas, or the electrical properties (e.g. relative permittivity, loss tangent, etc.) of the antenna environment. Finally, testing this approach on actual arrays and comparing the Actual (simulated) and Predicted beams with measured beams is essential.

Another area of future study is a more complete analysis of the sensitivity of the Predicted beam accuracy to errors in input parameters.  The fact that rather large errors in antenna positions (case B) did not degrade the RMSE significantly could be understood as arising because accurate S-parameters ``made up" for those errors.  However, it is then hard to understand why the RMSE is similar when errors are introduced to the S-parameters (case C).

Finally, in this paper we trained the DNN to recover the beam patterns of antenna arrays when operated as phased arrays with each dipole excited in phase with equal weights.  In future work we plan to explore cases with different excitations.  We also plan to train the DNN to recover the EEPs of each antenna in the array. The EEPs of each antenna are required for assigning excitation weights to synthesize array beams. EEPs are also required for radio interferometers, where the fundamental observables are the visibilities, which involve the products of the EEPs of the antenna pairs (Eq. \ref{eq:vis}).

\section{CONCLUSION}  
\label{sec:conc}

We have shown that a DNN can be trained to predict beam patterns of small phased arrays of radio antennas when given the design of the antennas, their operating frequency, their positions, and the S-parameters of the array network.  The predicted patterns are remarkably insensitive to errors in knowledge of the positions and the S-parameters. This work represents first steps toward applications where it is difficult or impossible to measure the beam patterns of an array directly and where electromagnetic simulations are limited by incomplete knowledge of the array properties or its environment, but where the S-parameters are known approximately. 

\section*{APPENDIX - Efficient S-parameter measurements}

For arrays with small numbers of antennas, S-parameters are typically measured using a vector network analyzer (VNA), connected sequentially between all the pairs of antenna ports in the array.  For an array of N antennas, this approach requires performing a sequence of N(N-1) measurements, a time-consuming process for large N.  In principle, one could realize an alternate approach where the array could measure its own S-parameters automatically.  Calibrated signals could be transmitted from each antenna and received at all the other antennas.  To do this, each antenna in the array could be outfitted with an ``embedded VNA (eVNA)" that, for example, could use a Radio Frequency System-on-Chip (RF-SoC) to transmit and receive the calibration signals.
The cadence of these measurements could be adjusted to accommodate the rate at which the array parameters might change due to antenna reconfiguration or environmental factors (e.g. temperature changes, foreign objects moving near the antennas, etc.).

\section*{ACKNOWLEDGMENT}
The authors would like to thank Moritz Muenchmeyer for advice on applying machine learning to this problem. We also thank Adam Beardsley and the anonymous referees for suggestions on improving the paper.



\bibliographystyle{IEEEtran}
\bibliography{references}

\end{document}